\documentclass[fleqn,10pt]{wlscirep}
\usepackage[utf8]{inputenc}
\usepackage[T1]{fontenc}
\usepackage{graphicx}
\usepackage{quantikz}
\usepackage{subcaption}
\usepackage{amsmath,amssymb,amsfonts}
\usepackage{multirow}
\usepackage{changes}
\usepackage{float} 
\usepackage{tikz}
\usepackage{quantikz}
\usepackage{enumitem}
\usepackage{cite}
\usepackage{algorithm}
\usepackage{algpseudocode}
\usepackage{booktabs}
\usepackage{makecell}
\usepackage{graphicx}
\usepackage{enumitem}
\usepackage{textcomp}
\usepackage{xcolor}

\title{\textcolor{black}{On the effects of emulated quantum layers on CNN performance and explainability for OCT and dementia imaging}}

\author[1,2]{Guillermo Rubiños Rodríguez}
\author[1]{Martín Ottavianelli}
\author[1]{Mateo Alonso}
\author[1]{Gonzalo Blázquez Gil}
\author[3,4]{Boris-Stephan Rauchmann} 
\author[1,+]{Pablo Díez-Valle}
\author[3,4,+,*]{Sergio Altares-López}
\affil[1]{Instituto Tecnológico de Galicia, Cantón Grande 9, Planta 3, 15003 A Coruña, Spain}
\affil[2]{Centro Singular de Investigación en Tecnoloxías Intelixentes (CiTIUS), Universidade de Santiago de Compostela, Santiago de Compostela, 15782, Spain}
\affil[3]{Department of Neuroradiology, LMU University Hospital, Ludwig Maximilian University of Munich, Marchioninistraße 15, 81377 Munich, Germany}
\affil[4]{Department of Psychiatry and Psychotherapy, LMU University Hospital, LMU Munich, Nussbaumstraße 7, 80336 Munich, Germany}
\affil[+]{These authors contributed equally to this work}

\affil[*]{slopez@med.lmu.de}
\newcommand{\acc}[2]{\makecell{$#1$\\
[-1pt]{\scriptsize$\pm #2$}}} 
\newcommand{\guille}[1]{{\textcolor{black}{#1}}}


\begin{abstract}
Hybrid quantum-classical convolutional neural networks have been proposed as an alternative to classical deep learning models. However, executing network components on quantum hardware imposes fundamental limitations, such as gradient estimation through repeated measurements, and increasing the number of qubits \guille{can lead} to trainability problems such as barren plateaus. Moreover, the reported success of these hybrid networks has been linked to quantum architectures that can be efficiently simulated on classical computers. Building on this observation, we investigate whether small, classically simulated quantum circuits can serve as useful components within larger models, as a substitute for purely classical layers. We present a systematic comparison between a Hybrid Quantum-inspired Convolutional Neural Network (HQiCNN) and \guille{standard Convolutional Neural Network (CNN) architectures}, where the models differ only in a single intermediate layer: an emulated quantum circuit in the HQiCNN and \guille{classical counterparts in the CNN}. \guille{The} models are evaluated on two medical imaging datasets (retinal OCT and brain MRI for dementia staging) while varying training set size, learning rate, batch size, circuit entanglement and observable set, so that the comparison does not depend on a single dataset or hyperparameter configuration. \guille{Under a strict parameter budget, the HQiCNN outperforms the classical baselines. When the classical layer is given more parameters, overall performance becomes comparable, with the HQiCNN showing improvements only in the intermediate-data regime.} Removing entanglement yields comparable performance while substantially improving the scalability of the simulations, and \guille{richer observable sets yield no performance improvement}. Finally, we introduce two SHAP-based metrics for comparing model explanations, $|$SHAP$|$IoU and EMD$_{\mathrm{pos}}$. \guille{They show that all architectures attend to anatomically plausible regions, and that the spatial agreement between their explanations increases with training set size.} These results indicate that, under specific data regimes, quantum-inspired hybrid models are a competitive alternative to classical CNNs for medical image classification.
\end{abstract}

\begin{document}

\flushbottom
\maketitle
    
\thispagestyle{empty}

\section{Introduction}
Quantum machine learning (QML)~\cite{qml_2017} has emerged as a promising paradigm at the intersection of quantum computing and artificial intelligence (AI), with growing interest in its application to domains where classical models face persistent limitations in accuracy, robustness, or interpretability. Among these domains, medical image diagnosis stands out as a particularly demanding use case: clinical decision support systems must not only achieve high predictive performance but also offer a degree of transparency that allows practitioners to trust and validate model outputs. Variational quantum algorithms (VQAs) \cite{cerezo2021variational}, and in particular parameterized quantum circuits (PQCs) \cite{benedetti2019parameterized} embedded within hybrid quantum-classical architectures, have been proposed as a means of introducing nonlinear, high-dimensional feature transformations that may complement existing machine learning (ML) approaches, including classical deep learning (DL) models \cite{PhysRevLett.127.090506, liu2021hybrid,long2025hybrid}.

% Despite this promise, the current state of the art in QML remains constrained by several well-documented challenges. Trainability issues such as barren plateaus, where gradients vanish exponentially with circuit depth or qubit count, have been shown to severely limit the scalability of PQC-based models \cite{cerezo2021costfunction}. Furthermore, recent benchmarking efforts have raised important questions regarding whether reported quantum advantages in QML are genuine or artifacts of insufficiently controlled experimental comparisons with classical baselines \cite{bowles2024betterclassicalsubtleart}. These concerns are compounded in near-term settings by hardware noise, limited qubit connectivity, and the substantial computational cost of simulating entangled quantum circuits classically, which together restrict the practical scale at which hybrid models can currently be evaluated.

Specifically in the context of QML for image classification, there have been several proposals such as introducing variational PQCs after RESNET18 frameworks for transfer learning\cite{Mari2020transferlearningin}, replacing the classical convolutional layer by PQCs \cite{long2025hybrid} and substituting the classical dense layers inside convolutional architectures by PQCs \cite{Senokosov2024}. However, most of these studies are constrained either by effective classical simulability \cite{qcnncerezo} or by well-documented challenges like trainability issues related to the limited qubit count and barren plateaus, which have been shown to severely limit the scalability of PQC-based models \cite{cerezo2021costfunction}. Furthermore, the results of the previous image-classification studies do not address the recently raised important questions regarding whether reported quantum advantages in QML are genuine or artifacts of insufficiently controlled experimental comparisons with classical baselines, and if quantum properties are even necessary for the reported advantages \cite{bowles2024betterclassicalsubtleart}. 

At the same time, the broader ML literature applied to medicine has increasingly emphasized the need for models that are not only accurate but also simple and interpretable. While architectures such as Vision Transformers (ViT)\cite{parvaiz2023vision} have achieved state-of-the-art performance on many imaging tasks, their complexity and limited transparency make them less suitable for clinical contexts, where explainability tools such as SHAP (SHapley Additive exPlanations) are important for justifying model predictions and building trust among medical practitioners \cite{lundberg2017unified}. This motivates the exploration of hybrid quantum-classical convolutional neural networks (HQCNNs) \cite{liu2021hybrid} as an alternative that retains the explainable AI and data efficiency of classical convolutional backbones while introducing a compact, parameter-matched quantum layer whose contribution to performance can be isolated and rigorously assessed.

% \textcolor{blue}{In this context, we propose a new framework that combines purely classical convolutional neural networks (CNN) \cite{lecun2015deep} and classically executed emulated PQCs (EQC), which we introduce as Hybrid Quantum-inspired Convolutional Neural Network (HQiCNN). We address the open challenge of a fair, architecturally controlled comparison to determine under which conditions the quantum layer provides a measurable advantage \cite{bowles2024betterclassicalsubtleart}} 
In this context, we propose a new framework that integrates classical convolutional neural networks (CNNs) \cite{lecun2015deep} with small-size classically emulated PQCs (EQCs), forming the Hybrid Quantum-inspired Convolutional Neural Network (HQiCNN). Our approach enables a fair, architecturally controlled comparison to investigate the conditions under which the integration of an emulated quantum layer provides measurable advantages \cite{bowles2024betterclassicalsubtleart}. To this end, we design two \guille{alternative classical} models that share the \guille{same convolutional backbone as the HQiCNN and differ only in the intermediate layer, an eight-parameter MLP in one model and a Dense layer in the other}, \guille{thereby allowing any observed performance improvements to be attributed to the integration of an emulated quantum layer in the architecture}. We evaluate the proposed architectures on two clinically relevant imaging tasks, retinal optical coherence tomography (OCT) \cite{kermany2018oct} classification and Alzheimer's disease staging from structural MRI (OASIS-1) \cite{marcus2007oasis}, across a broad hyperparameter space including training set size, learning rate, batch size, circuit entanglement, and observable set size. In addition, instead of directly comparing the model-dependent SHAP maps, we derive two SHAP-based methods that allow us to identify whether the models rely on comparable or divergent evidence when making predictions. The first tool is $|\text{SHAP}|IoU$, which plots \guille{, over a selected image,} the intersection areas between \guille{the selected top percentage of the} absolute SHAP maps of the models; while the second metric, $\text{EMD}_\text{pos}$, quantitatively compares the similarity between the positive SHAP distributions \guille{across a given set of images}. The combination of both tools allows us to observe that models rely on similar, \guille{possibly} relevant areas for a given image, while also showing consistent positive SHAP distributions across the whole test set.

The main contributions of this article are: (i) the exploration and proposal of emulated quantum circuits embedded within classical CNN architectures; (ii) a fair and comprehensive hyperparameter analysis to ensure an unbiased comparison between classical and quantum-emulated models; (iii) a systematic study of the impact of observables and entanglement on the performance of the quantum-emulated model; (iv) the evaluation of the robustness of this comparison across two challenging medical image classification datasets; and (v) the design and implementation of SHAP-based explainability methods to compare the models' behaviors.

This study is organized as follows. Section~\ref{sec:fundamentalsQC} introduces the fundamentals of quantum computing and hybrid parametric quantum-classical circuits. Section~\ref{sec:methods} describes the proposed models to be compared and the experimental methodology, including the description of the experiments conducted and the SHAP-based explainability tools used to compare results across models. Section~\ref{sec:results} presents the results and discussion, covering the effects of training set size, learning rate, batch size, entanglement, and observable set size, together with the explainability analysis based on SHAP attributions. Finally, Section~\ref{sec:conclusions} summarizes the main conclusions of this study and outlines directions for future work.
%Quantum machine learning (QML) applications to different areas of interest such as medical image diagnosis 
%\textcolor{blue}{Estructura: Intro QML, Estado del arte  - challenge (Schuld, BPs, Cerezo, VQAs), Motivación and innovation in the SOA (ML para medicina, menos complejo, mas preciso, mas intepretable (ver shap), y nuestra solución brevemente.
%+ Estructura del artículo.}

\section{Fundamentals of Quantum Computing}
\label{sec:fundamentalsQC}
Quantum computing is based on the principles of quantum mechanics, with the \textit{qubit} as its fundamental unit. A qubit is described by a vector in a two-dimensional complex Hilbert space and can exist in a linear combination of the basis states $|0\rangle$ and $|1\rangle$, a property known as \textit{superposition}:
\[
|\psi\rangle = \alpha|0\rangle + \beta|1\rangle = \begin{bmatrix}
\alpha \\
\beta
\end{bmatrix},
\] 
where $\alpha$ and $\beta$ are complex amplitudes satisfying $|\alpha|^2 + |\beta|^2 = 1$ \cite{nielsen2010quantum}. Entanglement is a quantum correlation between subsystems such that the joint state cannot be factorized into a tensor product of individual states. Entangled qubits allow quantum circuits to encode joint probability distributions and correlations that are difficult to represent classically.

Quantum computations are implemented using \textit{quantum gates}, which are unitary operators acting on one or more qubits. Single-qubit rotation gates around the Bloch sphere axes are commonly used in hybrid models:
\begin{equation}
\begin{aligned}
R_X(\theta) &=
\begin{bmatrix}
\cos\left(\frac{\theta}{2}\right) & -i\sin\left(\frac{\theta}{2}\right) \\
-i\sin\left(\frac{\theta}{2}\right) & \cos\left(\frac{\theta}{2}\right)
\end{bmatrix}, \quad %\\[0.5em]
R_Y(\theta) &=
\begin{bmatrix}
\cos\left(\frac{\theta}{2}\right) & -\sin\left(\frac{\theta}{2}\right) \\
\sin\left(\frac{\theta}{2}\right) & \cos\left(\frac{\theta}{2}\right)
\end{bmatrix}, \quad%\\[0.5em]
R_Z(\theta) &=
\begin{bmatrix}
e^{-i\theta/2} & 0 \\
0 & e^{i\theta/2}
\end{bmatrix}.
\end{aligned}
\end{equation}

These gates rotate a qubit state around the corresponding axis of the Bloch sphere. Multi-qubit gates, such as the CNOT, create entanglement by flipping a target qubit conditioned on the control qubit. In digital quantum computing, the most common observables used are different tensor product combinations of Pauli matrices~\cite{nielsen2010quantum}:
\begin{equation}
\begin{aligned}
X &=
\begin{bmatrix}
0 & 1 \\
1 & 0
\end{bmatrix}, \qquad \qquad \qquad \qquad \qquad%\\[0.5em]
Y &=
\begin{bmatrix}
0 & -i \\
i & 0 
\end{bmatrix}, \qquad \qquad \qquad \qquad \qquad%\\[0.5em]
Z &=
\begin{bmatrix}
1 & 0 \\
0 & -1
\end{bmatrix}.
\end{aligned}
\end{equation}

The sequencing of single-qubit rotations with tunable parameters $\mathbf{\theta}$ defines a PQC, as shown in Figure \ref{fig:architectures}\textcolor{blue}{b}. PQCs act as trainable nonlinear transformations of input data in hybrid quantum-classical models and form the basic building block of so-called quantum neural networks (QNNs) \cite{abbas2021power}. A high-dimensional representation of the input data is obtained after the parameterized evolution, from which classical features are extracted by estimating the expectation values of a selected set of observables through repeated sampling (shot-based measurement) of the quantum state; these classical outputs can then be processed classically to predict or classify. The parameters $\mathbf{\theta}$ are optimized using classical gradient-based methods, with the gradient conventionally computed via the parameter-shift rule, which yields exact analytical gradients by evaluating the circuit at shifted parameter values and is compatible with optimizers commonly used in deep learning. Notable limitations include barren plateaus \cite{cerezo2021costfunction}, which hinder training as gradients vanish exponentially with circuit size \cite{mcclean2018barren}, the noise inherent to current quantum hardware \cite{Wang_2021}, and the finite sampling limitation.
\section{Methods}
\label{sec:methods}
In this section, we introduce the proposed models that are to be compared on health-related image classification tasks. Then, we describe the experimental setup that allows us to fairly compare the algorithms. Finally, we introduce the two SHAP-based tools derived to compare and explain the predictions provided by the models.
\subsection{Quantum-inspired and classical arquitectures}
\label{sec:architectures}
\begin{figure}[t]
    \centering
    \includegraphics[width=\linewidth]{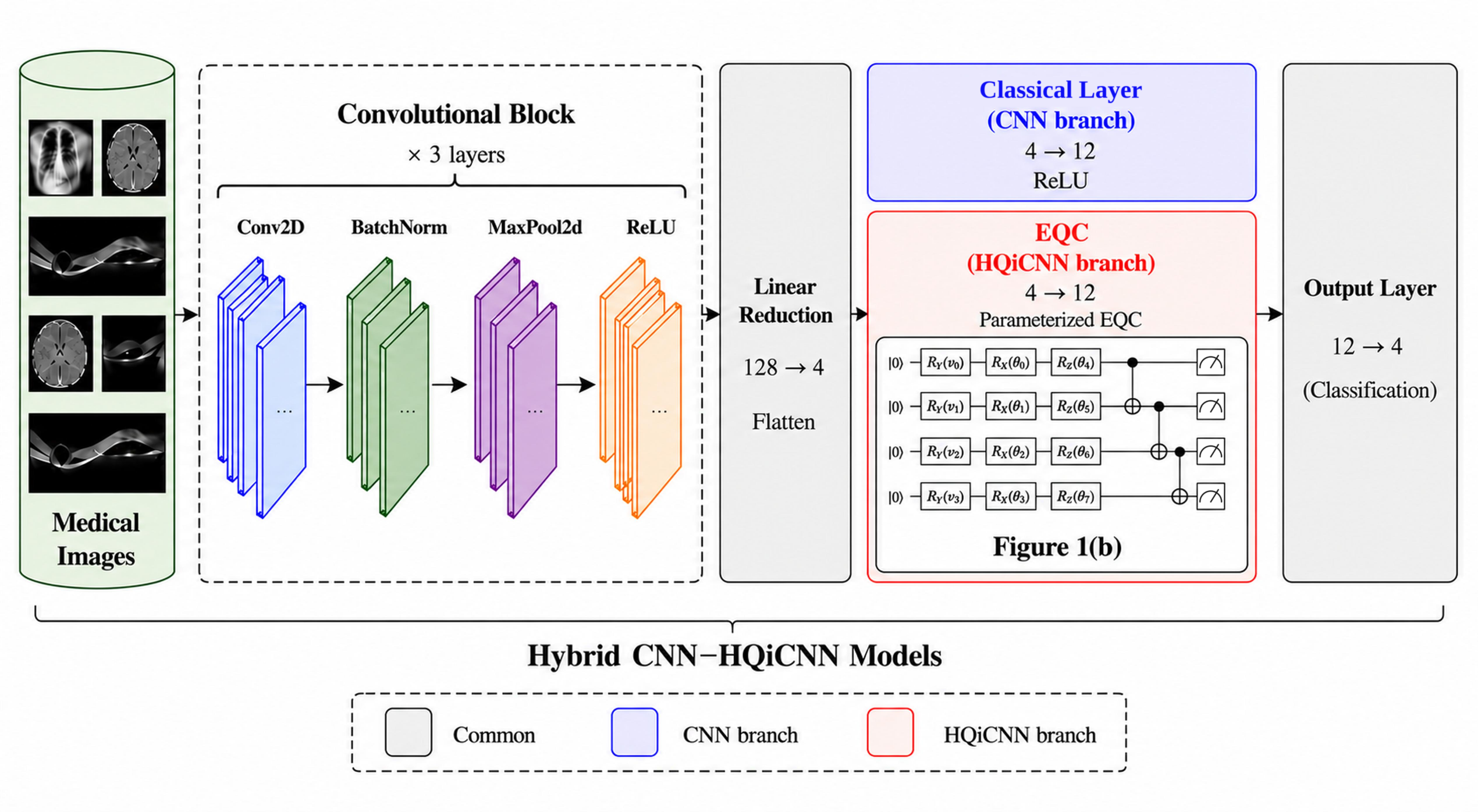}
    \caption{Schematic representation of the model to compare. \guille{The shared} common architecture and only differ in the layer right before the output. The HQiCNN uses a preprocessing scaled sigmoid transformation to map the input data into the quantum circuit. \guille{The classical layer is an MLP8 for the CNN-MLP8 model and a Dense layer for the CNN-Dense architecture.} \textbf{(b)} Parameterized EQC used within the HQiCNN algorithm.}
    \label{fig:architectures}
\end{figure}

The models we compare in this article are schematized in Figure \ref{fig:architectures}. \guille{All of them} share a common classical backbone that comprises three main components. First, a convolutional block composed of three convolutional layers, each incorporating ReLU activation, a kernel size of three, batch normalization, and max pooling, transforms the three-dimensional \textit{RGB} input into a 128-dimensional feature vector. This vector is then compressed by a linear reduction layer, a linear layer without an activation function that reduces the 128-dimensional vector into a 4-dimensional vector, which can be encoded into either the quantum circuit or the \guille{alternative} layer of the CNN branch. Finally, an output layer, also a linear layer without an activation function, takes the 12-dimensional vector from the previous layer and maps it to a vector whose size corresponds to the number of labels.

\guille{The proposed architectures differ in a single layer, which allows the effect of the emulated quantum circuit to be isolated and compared directly with that of alternative classical ReLU-activated layers. The components of each model are described in the following subsections, and Table \ref{tab:parameters} reports, for the three models, the total number of trainable parameters as well as those associated with the layer under study.}

%The models we compare in this article are schematized in Figure \ref{fig:architectures}. Both models share a common classical backbone. The common layers are:
%\begin{itemize}
 %   \item \textbf{Convolution Block} of 3 convolutional layers. They have ReLu activation, a kernel size of 3, BatchNormalization and MaxPool2d. They transform the \textit{RGB} 3 dimensional input into a 128 dimension vector.
  %  \item \textbf{Linear Reduction layer}. It consists in a linear layer with no activation function that reduces the 128 dimension vector into a 4 dimension vector that we can encode into both the Quantum Circuit and the CNN branch linear layer.
    %\item \textbf{Output Layer}. It is a linear layer with no activation function that takes the 12 vector from the previous layer into a vector whose size is the number of labels.
%\end{itemize}
%The proposed architectures differ only in a single layer, allowing for a fair and accurate comparison of the impact of the simulated quantum circuit relative to a classical ReLu activated neural network. We carefully described the different components in the next subsections.
\subsubsection{\guille{Classical baselines (CNN branch)}}
\guille{We compare the HQiCNN against two classical baselines, each obtained by replacing the quantum circuit with a classical layer that receives the same 4-dimensional feature vector from the Linear Reduction layer and produces a 12-dimensional output. Each baseline controls for a different source of potential bias. The first, CNN-MLP8, is a small multilayer perceptron with exactly 8 trainable parameters, matching those of the quantum circuit, followed by a linear layer that maps its output to 12 dimensions. Since both models have the same trainable capacity in the layer under study, any difference in performance can be attributed to the type of transformation rather than to the number of parameters. The second, CNN-Dense, is a standard fully connected layer from 4 to 12 dimensions, which is the conventional choice in classical CNNs at this position. This baseline represents the most direct classical replacement for the quantum circuit, without restricting its parameter count.} This combination of convolutional and linear layers follows the standard design of deep convolutional models, providing a simple transition between feature extraction and the subsequent processing stages.

We choose a convolutional backbone over transformer-based architectures such as the Vision Transformer (ViT) for the following reasons. CNN-based models remain simpler and more interpretable than transformer-based architectures, a property of particular relevance in medical applications, where model transparency and the ability to justify predictions are often as important as raw performance. They also typically require fewer data and computational resources to train, resulting in faster convergence, which is advantageous given the moderate size of the dataset used in this study. Most importantly, the CNN branch is designed to serve as a fair classical baseline against which the quantum branch can be meaningfully compared, since both architectures share the same backbone and a comparable number of trainable parameters, differing only in the layer under study. Introducing a substantially more complex model, such as a ViT, would compromise this parameter-matched, architecture-controlled comparison, making it difficult to attribute performance differences specifically to the quantum component rather than to differences in model capacity or design.

\subsubsection{Emulated Quantum Circuit (HQiCNN branch)}
The quantum-inspired branch replaces the dense neural layer \guille{of the CNN-Dense model} with the parameterized emulated quantum circuit (EQC) shown in Figure~\ref{fig:architectures}\textcolor{blue}{b}. The distinctive feature of emulated quantum circuits is that, whether due to their low correlations or their small size, they can be efficiently computed on classical devices. Since we are considering quantum architectures with only four qubits, they can be computed easily by quantum hardware emulation within a short time.  

The four-dimensional feature vector produced by the Linear Reduction layer is first encoded into the quantum state through angle encoding, using $R_Y$ rotations applied to each qubit. A variational layer then applies trainable $R_X$ and $R_Z$ rotations, followed by a linear entanglement layer implemented with CNOT gates. Finally, the three local Pauli observables ($X$, $Y$, and $Z$) are measured on each of the four qubits, yielding a $12$-dimensional output vector that is passed to the output classical layer.\par
Since parameterized quantum gates implement rotations and therefore require input angles bounded within $\left[0, 2\pi\right)$, we investigate how the choice of preprocessing applied to the four-dimensional vector $\mathbf{v}$ prior to encoding affects downstream performance, comparing three distinct strategies. The first employs a scaled sigmoid transformation, $2\pi \cdot \text{sigmoid}(\mathbf{v})$, which smoothly maps arbitrary real-valued inputs into the required interval while compressing extreme values toward its boundaries. The second applies a modulo transformation, $\text{remainder}(\mathbf{v}, 2\pi)$, which wraps the input periodically into the valid range without altering its relative scale, thereby preserving the original distances between values more faithfully than the sigmoid mapping. The third condition serves as a baseline in which no transformation is applied and the raw vector $\mathbf{v}$ is encoded directly, allowing us to assess whether explicit preprocessing offers any measurable \guille{improvement} over leaving the input unconstrained. A study comparing these three options showed that the scaled sigmoid transformation consistently outperforms the other ones, as a result, we select this transformation for the rest of the study.

Finally, we highlight that the proposed quantum-inspired implementation allows the hybrid model to compute gradients analytically via backpropagation, circumventing the need to estimate the quantum gradient through the parameter shift rule~\cite{paramshiftrule}. The parameter shift rule requires two forward passes per parameter to estimate the gradient, resulting in a total of $2p$ forward passes for a quantum circuit of $p$ parameters. This quantum-inspired architecture can implement backpropagation to calculate all gradients in parallel using one single backward pass, effectively reducing the number of passes required to calculate the gradient from $O(p)$ to one.

\begin{table}[t]
\centering
\caption{Comparison of total trainable parameter counts and trainable parameter counts in the studied layer between HQiCNN and the classical baselines.}
\label{tab:parameters}
\begin{tabular}{lcc}
\toprule
 & \textbf{Total Parameters} & \textbf{Trainable Parameters in the Studied Layer} \\
\midrule
CNN-MLP8  & 110884 & 8  \\
CNN-Dense & 110876 & 60 \\
HQiCNN    & 110832 & 8  \\
\bottomrule
\end{tabular}
\end{table}
\subsection{Experimental setup}
In order to ensure a faithful comparison between the results obtained from \guille{the considered} architectures and to examine how the quantum layer affects \guille{the overall performance of the model}, we study the models' hyperparameters as explained in Ref. \cite{bowles2024betterclassicalsubtleart}. We compare the accuracy of the classical and quantum emulated models for the OASIS and OCT datasets (see Section \ref{sec:datasets}), ensuring that the effects are not dataset-specific. We further examine the role of the hybrid quantum layer by evaluating the effect of entanglement on performance. \guille{To ensure statistical robustness, each experiment is repeated with ten random seeds and the results are reported for each training set size as the mean ± the half-width of the 95\% confidence interval.} Table \ref{tab:hyperparam_search} displays the studied hyperparameters for the OCT dataset. \guille{We report the validation accuracy results for the OCT dataset and choose hyperparameter configurations to transfer to the OASIS dataset based on these results. Furthermore, we include the test results in the Supplementary Information.} Since the balanced OASIS dataset was considerably smaller compared with the OCT dataset, we used the OASIS dataset experiment to help us determine the robustness of the model, assessing whether the observed trends among the models generalize across different datasets. Note that this robustness test is performed in a simplified 2D setting (see Section \ref{sec:datasets}) and is not intended to provide clinical conclusions, but rather serves as a robustness check, evaluating whether the behavior of the architectures generalizes to a distinct imaging modality and classification task. \par
\begin{table}[ht]
\centering
\caption{\guille{The complete hyperparameter search space used on the OCT dataset is described.} The impact of the learning rate and training size combination on model performance is studied \guille{for three models}, resulting in \guille{810} executions. Subsequently, the best learning rate for each model at a training size of 1000 samples is used to investigate the effect of batch size, resulting in \guille{90} additional executions. Finally, the quantum properties of the HQiCNN are further analyzed by evaluating the importance of entanglement and the observable set across different training sizes, adding a total of 340 executions.}
\label{tab:hyperparam_search}
% \resizebox{0.85\textwidth}{!}{%
\begin{tabular}{lcc}
\toprule
\textbf{Hyperparameter} & \textbf{Values} & \textbf{\# Values} \\
\midrule
Models          & $\{\text{HQiCNN, CNN-MLP8, CNN-Dense}\}$          & 3 \\
Training Size   & $\{200,400,800,1000,1200,1600,2000,12000,30000\}$ & 9 \\
Learning Rate   & $\{10^{-2},10^{-3},10^{-4}\}$                     & 3 \\
Batch Size      & $\{8,16,32\}$                                     & 3 \\
Quantum Circuit & \{Entangled, Not Entangled\}                      & 2 \\
Observable Set  & $\{\mathcal{O}_1,\mathcal{O}_2,\mathcal{O}_3\}$   & 3 \\
Seeds           & 10 randomly selected seeds                        & 10 \\
\midrule
\textbf{Total configurations} & \multicolumn{2}{c}{\textbf{$9\times3\times10\times3 + 3 \times 10 \times 3 + 2 \times 8 \times 10 + 3 \times 6\times 10=1240$}} \\
\bottomrule
\end{tabular}%
% }
\end{table}
% \begin{table}[ht]
% \centering
% \caption{The hyperparameter search space used for both models evaluated on the OCT dataset is described. The impact of the learning rate and training size combination on model performance is studied, resulting in 270 executions. Subsequently, the best learning rate for each model at a training size of 1000 samples is used to investigate the effect of batch size, resulting in 30 additional executions. Finally, the quantum properties of the HQiCNN are further analyzed by evaluating the importance of entanglement and the observable set across different training sizes, adding a total of 340 executions.}
% \label{tab:hyperparam_search}
% \resizebox{\textwidth}{!}{%
% \begin{tabular}{|ll|c|}
% \hline
% \multicolumn{1}{|l|}{\textbf{Hyperparameter}} & \textbf{Values} & \textbf{\# Values} \\ \hline
% \multicolumn{1}{|l|}{Training Size} & $\{200,400,800,1000,1200,1600,2000,12000,30000\}$ & 9 \\ \hline
% \multicolumn{1}{|l|}{Learning Rate} & $\{10^{-2},10^{-3},10^{-4}\}$ & 3 \\ \hline
% \multicolumn{1}{|l|}{Batch Size} & $\{8,16,32\}$ & 3 \\ \hline
% \multicolumn{1}{|l|}{Quantum Circuit} & \{Entangled, Not Entangled\} & 2 \\ \hline
% \multicolumn{1}{|l|}{Observable Set} & $\{\mathcal{O}_1,\mathcal{O}_2,\mathcal{O}_3\}$ & 3 \\ \hline
% \multicolumn{1}{|l|}{Seeds} & 10 randomly selected seeds & 10 \\ \hline
% \multicolumn{2}{|l|}{\textbf{Total configurations tested}} & \textbf{$9\times3\times10 + 3 \times 10 + 2 \times 8 \times 10 + 3 \times 6\times 10=640$} \\ \hline
% \end{tabular}%
% }
% \end{table}

The remaining hyperparameters, those not included in Table \ref{tab:hyperparam_search}, related to the training of the models were kept constant across all realized experiments. We highlight the use of \textit{GradScaler}, which enables the optimizer to handle the quantum and classical gradients in a unified manner. Given the multiclass nature of the dataset, cross-entropy is selected as the loss function. This choice also justifies the absence of a softmax activation in the output layer, since the cross-entropy loss operates directly on unnormalized logits.\par

\guille{We provide in Table S1 (see Supplementary Information) all the necessary optimization information and training setup for ensuring the reproducibility of the results presented in this article. All models are implemented using PyTorch \cite{pytorch}, including the quantum component, which is simulated through native PyTorch functions rather than a dedicated quantum computing framework.}
% \textcolor{blue}{Maybe here a table with the seeds tested and the training sizes. Also, other training hyperparameters such as batch size, starting lr, number max epochs  (also why - make it standard, based on the previous literature not necessary more etc..)...}

% \textcolor{blue}{Maybe also include why we tested all these hyperparameters and cite Maria Schuld article on QML models benchmarking \cite{bowles2024betterclassicalsubtleart}.}
%\subsection{Implementation}

% The parameter shift rule requires two forward passes per parameter to estimate the gradient, resulting in a total of $2p$ forward passes for a quantum circuit of $p$ parameters \cite{paramshiftrule}. In contrast, the PyTorch Autograd function naturally implements backpropagation to calculate all gradients in parallel using a single backward pass. This PyTorch quantum-inspired implementation reduces the number of passes required to calculate the gradient from $O(p)$ to one.
% \subsection{Experiments}
% \textcolor{blue}{Incluir: Test Acc vs tr size for CNN and HQCNN with and without entanglement (argue why the selection of observables for this current study (local))}
\subsection{Explainability tools}

\guille{The interpretability of deep neural networks is inherently limited, as their decision rules are distributed across many hidden layers and a large number of trainable parameters.} Several families of explainability techniques have been proposed to address this challenge. Gradient-based visualization methods such as Grad-CAM~\cite{selvaraju2017grad} and its refinement Grad-CAM++~\cite{chattopadhay2018grad}, generate class-discriminative localization maps by leveraging the gradients flowing into the final convolutional layers, offering an intuitive visualization of the image regions driving a CNNs prediction; however, these approaches are architecture-dependent, typically restricted to convolutional backbones, and provide coarser, lower-resolution attributions than perturbation-based alternatives. Attention-based methods, applicable to transformer architectures, and causal interpretability approaches, which attempt to move beyond mere correlation by estimating the causal effect of specific input features or latent factors on the model output~\cite{moraffah2020causal}, represent complementary directions, though they generally demand additional modeling assumptions or architecture-specific access.

In contrast, perturbation-based, model-agnostic tools such as SHAP (SHapley Additive exPlanations)\cite{lundberg2017unified} or LIME (Local Interpretable Model-Agnostic Explanations)~\cite{ribeiro2016should} are widely adopted in the field of AI applied to medicine, as they can be applied uniformly across the different model architectures compared in this study and provide evidence that ML models are attending to anatomically relevant areas, rather than relying on spurious or irrelevant features~\cite{vimbi2024interpreting}. We therefore adopt SHAP as the basis for our interpretability analysis and propose two complementary techniques built upon it: the first is a tool that compares, for a single image, the SHAP areas that contribute the most to the predictions of the different models; the second is a metric that quantifies, across the whole test set, the similarity between the positive SHAP maps generated by these models.

% Interpreting the results obtained for ML algorithms comprised of hidden layers and large trainable parameter counts is a difficult task. The use of perturbation-based tools such as SHAP or LIME~\cite{ribeiro2016should} is \textcolor{blue}{widely adopted} in the field of AI applied to medicine in order to \textcolor{blue}{provide evidence that ML models are attending to anatomically relevant areas, rather than relying on spurious or irrelevant features}~\cite{vimbi2024interpreting}. We propose two interpretability techniques: the first is a tool that compares, for a single image, the SHAP areas that contribute the most to the prediction of both models; the second is a metric that quantifies, across the whole test set, the similarity between the positive SHAP maps generated by both models.

% Interpreting the results obtained for ML algorithms comprised of hidden layers and large trainable parameter counts is a difficult task. The use of perturbation-based tools such as SHAP or LIME~\cite{ribeiro2016should} is standard in the field of medical image classification to validate that the ML models are studying areas of medical interest~\cite{vimbi2024interpreting}. We propose two interpretability techniques: the first is a tool that compares the SHAP areas that affect the most for the prediction for both models, and the second is a metric that quantifies the similarity between positive SHAP maps on average across the test images.

\subsubsection{Absolute SHAP Intersection of Union ($\text{|SHAP|}_{\text{IoU}}$)} \label{sec:SHAPiou_explain}

In order to effectively compare and validate that the models study the areas of interest for our images, we derive the $|\mathrm{SHAP}|_{\mathrm{IoU}}$. \guille{This is a visual explainability tool that compares the absolute SHAP maps for a given image between two models, allowing us to identify the areas of interest of each model ($R_A$ and $R_B$ respectively) for the same studied image. In our experiments, the results are averaged over ten randomly selected seeds. The absolute SHAP map of each seed is normalized and the resulting maps are summed, so that all seeds contribute equally to the aggregated map. The final aggregated maps are then overlaid on the original image, indicating the regions each model concentrates its attention on independently as well as those shared. Since these aggregated maps cover the whole image, we only plot the regions for the top 2\% and 10\% highest absolute SHAP values to study where the different models pay more attention when predicting. These percentages are selected by comparing the number of shared pixels against the total pixel count of the image across thresholds. The supporting plots and results obtained using different percentages are reported in Section S5 of the Supplementary Information.}

% In order to effectively compare and validate that both models are studying the medical areas of interest for our datasets, we derived the |SHAP|IoU. This tool takes the top $10\%$ absolute SHAP values of the image distribution pixels for each model, then plots them over the taken image highlighting the intersection and single model interest areas. This tool enables us to certify which areas the models intersect, proving similarity between them, and in which areas they differ, identifying interpretable image based medical explanations on why one model performs better than the other one.

\subsubsection{Positive SHAP Earth Mover's Distance (\texorpdfstring{EMD$_{pos}$}{EMD_pos})}
\label{sec:emdpos}
The previous method serves as an interpretability tool for a single image at a time. With the aim of correctly comparing the similarity between the positive SHAP distributions \cite{lundberg2017unified} (areas that improve the correct prediction of the models) across several images, we introduce the metric $\text{EMD}_{pos}$. This metric is a generalization to two dimensions of the Wasserstein metric for one-dimension probabilistic distributions.

\guille{First, for each image and for each model trained with a different random seed, we normalize the positive SHAP values so that they form a two-dimensional probability distribution over the pixel grid. We then compute the pairwise Euclidean cost matrix between all pixel coordinates:}

\begin{equation}
    \text{EMD}_{\text{pos}} = min_{\gamma \in \prod(\omega_{HQiCNN}|\omega_{CNN})}\sum_{i,j}\gamma_{i,j}||p_i - p_j||_2.
\end{equation}
The $\omega_{HQiCNN | CNN}$ represents the normalized positive SHAP distribution for our models, and $p_{i,j}$ are the normalized coordinates of the pixels $p = \left(\frac{y}{64},\frac{x}{64}\right)$, and $\gamma_{i,j}$ specifies the amount amount of probability mass transported from location \textit{i} to target location \textit{j}. In our considered datasets, the images that we use are transformed to tensors of resolution $64\times64$ pixels to input into the model, so we explore $4096$ grid points.

This proposed metric has a lower and an upper bound. The lower bound indicates identical distributions while the upper bound indicates non correlated or opposite pixel distribution. The lower bound can be derived directly as $0$, while the upper bound depends on the resolution of the image. We derive the maximum bound by considering two distributions, one centered in the corner $p=\left(0,0\right)$ and the other distribution in the opposite corner position $p=\left(63/64,63/64\right)$.
\begin{equation}
    D = \sqrt{\left(\frac{63}{64}\right)^2 + \left(\frac{63}{64}\right)^2} = \frac{63}{64}\sqrt{2} \approx 1.3921.
\end{equation}
Therefore,
\begin{equation}
    0 \;\le\; \text{EMD}_{\text{pos}} \;\le\; \frac{63}{64}\sqrt{2} \approx 1.3921.
\end{equation}
% \added{We can multiply the resulting $\text{EMD}_{\text{pos}}$ by $64$ in order to get the pixel averaged displacement between positive SHAP attribution maps.}

\section{Results and Discussion}
\label{sec:results}

This section collects the discussion over the obtained results for the considered experiments that allow us to fairly compare the proposed architectures. First, we explain the considered datasets and their characteristics. Then, we discuss the hyperparameter results for the more balanced OCT dataset. Afterwards, we conduct a robustness test using the dementia dataset. Finally, a comparative study of explainability is performed for the OCT dataset.

\subsection{Datasets} \label{sec:datasets}

\subsubsection{Retinal optical coherence tomography (OCT)}

The dataset used in this study consists of retinal optical coherence tomography (OCT) images, a widely used imaging modality that provides high-resolution cross-sectional views of the retina and is routinely employed in ophthalmic diagnosis \cite{kermany2018identifying}. It includes 84,495 JPEG images organized into two standard splits (training and test) and four clinically relevant classes: NORMAL, CNV (choroidal neovascularization), DME (diabetic macular edema), and DRUSEN, as can be shown in Figures \ref{fig:drusenimg}, \ref{fig:healthycontrolimg}, \ref{fig:cnvimg}, \ref{fig:dmeimg}. Each image is labeled according to its diagnostic category and associated with an anonymized patient identifier and acquisition index to ensure traceability without compromising patient privacy. Data were collected from multiple international clinical centers between 2013 and 2017 using Spectralis OCT devices (Heidelberg Engineering). To ensure label quality, a multi-stage review process was applied. Images first underwent basic quality control to remove those with severe artifacts, followed by independent grading by multiple ophthalmologists \cite{kermany2018oct}. \guille{Based on the OCT images we resized to $64 \times 64$ pixels}. Final labels were confirmed by senior retinal specialists with extensive clinical experience. \guille{A detailed explanation of the training, validation and test set generation can be found in Sec.~\ref{trteva}}
% In addition, a subset of validation images was independently re-annotated to further assess and mitigate potential labeling inconsistencies \guille{(more details on Sec.~\ref{trteva})}.

\subsubsection{Alzheimer's Disease}\label{sec:OASIS_intro}

In order to test the robustness of the proposed frameworks across different datasets, we also apply it to a dementia staging dataset. We conducted experiments using the OASIS-1 MRI dataset, which contains T1-weighted structural brain images of $416$ subjects aged 18 to 96 years \cite{marcus2007oasis}. The original 3D image files were converted to Nifti format using the FSL tool, and later converted to images in the dataset preparation. Specifically, the images are first reoriented to a standard anatomical orientation and corrected for scanner-induced intensity inhomogeneities using a bias field correction algorithm. The field of view is then cropped to a standard brain size, and non-brain tissue (skull, scalp) is removed via skull-stripping. The resulting brain-extracted volumes are linearly registered to the MNI152 template, a standard stereotactic reference space of the human brain widely used in neuroimaging (Montreal Neurological Institute, 152-subject average) \cite{fonov2009unbiased,fonov2011unbiased}, at two~mm resolution, using a 12 degrees-of-freedom affine transformation with trilinear interpolation. Tissue segmentation is subsequently performed on the registered T1-weighted images, and a White Matter (WM) mask is obtained by thresholding the WM partial volume estimation map. This mask is applied to the original T1w image to isolate WM voxel intensities, and their mean value is computed. Each T1w volume is then intensity-normalized by dividing it by this WM mean value, following the Shinohara normalization \cite{shinohara2014statistical} approach, in order to reduce inter-subject and inter-scanner intensity variability. Finally, the resulting normalized 3D volumes are converted into 2D representations by extracting axial slices from each subject's registered brain volume, which are subsequently exported as images for use in the classification framework. \guille{Based on OASIS images we resized to $128 \times 128 $ pixels.}

\begin{figure}[ht]
    \centering
    % Fila 1
    \begin{subfigure}{0.45\textwidth}
        \centering
        \includegraphics[width=\linewidth]{images/dataset_example_images/DRUSEN-821543-1.jpeg}
        \caption{Drusen class sample.}
        \label{fig:drusenimg}
    \end{subfigure}
    \hfill
    \begin{subfigure}{0.45\textwidth}
        \centering
        \includegraphics[width=\linewidth]{images/dataset_example_images/NORMAL-2819870-1.jpeg}
        \caption{Healthy Control class sample.}
        \label{fig:healthycontrolimg}
    \end{subfigure}

    \vspace{0.5cm}

    % Fila 2
    \begin{subfigure}{0.45\textwidth}
        \centering
        \includegraphics[width=\linewidth]{images/dataset_example_images/CNV-235517-1.jpeg}
        \caption{Choroidal Neovascularization (CNV) class sample.}
        \label{fig:cnvimg}
    \end{subfigure}
    \hfill
    \begin{subfigure}{0.45\textwidth}
        \centering
        \includegraphics[width=\linewidth]{images/dataset_example_images/DME-145590-3.jpeg}
        \caption{Diabetic Macular Edema (DME) class sample.}
        \label{fig:dmeimg}
    \end{subfigure}
    
    \vspace{6mm}
    \begin{subfigure}{0.45\textwidth}
        \centering
        \includegraphics[width=0.93\linewidth]{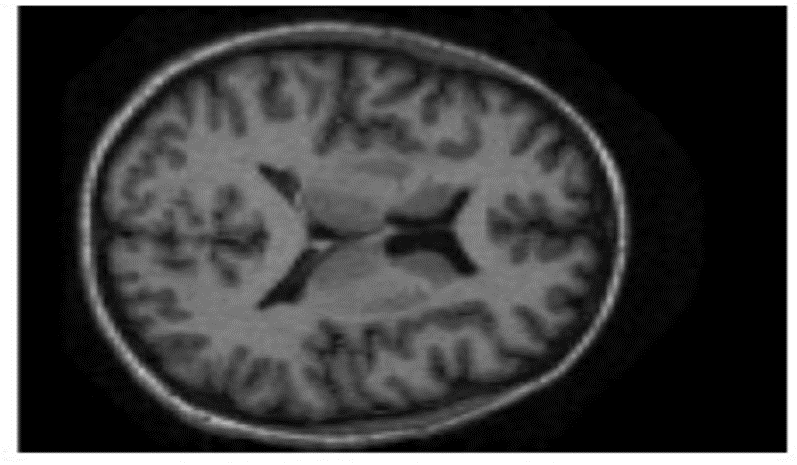}
        \caption{Non Demented class sample.}
        \label{fig:nondemenimg}
    \end{subfigure}
    \hfill
    % \begin{subfigure}{0.33\textwidth}
    %     \centering
    %     \includegraphics[width=0.97\linewidth]{images/very_mild_demen.png}   \caption{Very Mild Demential class sample.}
    %     \label{fig:verymilddemenimg}
    % \end{subfigure}
    % \hfill
    \begin{subfigure}{0.45\textwidth}
        \centering
        \includegraphics[width=\linewidth]{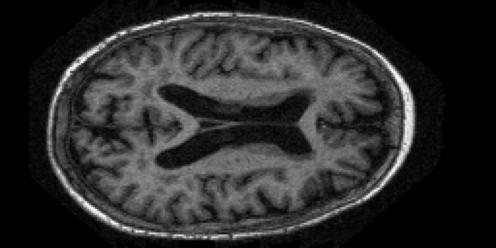}
        \caption{Mild/Moderate Dementia class sample.}
        \label{fig:milddemenimg}
    \end{subfigure}

    \caption{Class samples of the considered datasets. Sub-figures a, b,c  and d correspond to the retinal OCT images \cite{kermany2018identifying}. Sub-figures e, f and g correspond to the OASIS-1 Dementia dataset \cite{marcus2007oasis}. }
    \label{disea_eye}
\end{figure}

\guille{The OASIS dataset used in this study provides four cognitive stage labels: Non Demented, Very Mild Dementia, Mild Dementia, and Moderate Dementia. However, the dataset contains only 416 unique subjects, each contributing multiple MRI slices. To prevent subject-level data leakage, all images from a given subject must be assigned to the same partition (training, validation, or test), which leaves too few subjects in the less represented stages to build reliable and balanced subsets. For this reason, the original labels were regrouped into two classes: Non Demented and Demented, where the latter combines the Very Mild Dementia, Mild Dementia, and Moderate Dementia categories. Figures \ref{fig:nondemenimg} and \ref{fig:milddemenimg} show representative samples of each class. This binary formulation provides a sufficient number of subjects per class to evaluate the proposed HQiCNN framework under a controlled classification problem. Additionally, the majority class was downsampled to match the size of the minority class, ensuring a balanced class distribution and preventing the model from being biased towards the most represented class. The subject-level partitioning into training, validation, and test sets is described in Section \ref{trteva}.}

\subsubsection{Input resolution and normalization}

\guille{Images from the considered dataset datasets were spatially downscaled and their intensities normalized to the range $[0,1]$, as described above. This downscaling was deliberately adopted to reduce the computational cost and memory requirements of the simulations, which is particularly critical for the quantum-inspired and hybrid quantum--classical models, whose cost grows rapidly with input size. Low-resolution inputs also enable faster experimentation and allow the feasibility of the HQiCNN architectures to be assessed under the constraints of current quantum simulators, while preserving the coarse anatomical patterns required for classification.}

% \textcolor{blue}{Spatial downscaling was deliberately performed to reduce computational complexity and memory requirements, particularly for quantum-inspired and hybrid quantum–classical models with strict input-size constraints. Moreover, using low-resolution inputs facilitates faster experimentation. It allows testing the feasibility of HQiCNN architectures on resource-limited quantum hardware simulators while preserving the coarse anatomical patterns needed for early-stage classification.}

\subsubsection{\guille{Training, validation and test set partitioning}}
\label{trteva}
\guille{The OCT dataset provides separate training and test sets by design, which guarantees that no data leakage occurs between them. The validation set was obtained by randomly sampling 10\% of the training data used in the experiments. In contrast, the OASIS dataset \cite{marcus2007oasis} does not include a predefined subject-level partition. Since several 2D slices are extracted from each subject, a random split at the slice level could place slices from the same subject in different sets, leading to data leakage. To prevent this, the slices were first grouped by subject, and the subjects were then assigned to the training, validation and test sets, so that all slices from a given subject belong to a single set.}

\guille{In OASIS, every slice carries a subject identifier, so all sessions, scans and slices of the same brain share the same ID. We used this identifier to split the data at the subject level, so that no brain contributes slices to more than one of the training, validation and test sets. The resulting dataset, after downsampling the larger Non Demented class to match the number of subjects, contains 81 subjects for each of the two classes, Demented and Non Demented, with 15029 and 15761 slices available, respectively. Because the Demented class merges several of the original OASIS severity levels, subjects were assigned to each set stratified by severity level, so that every level is represented at least in the training and test sets. The test set was built first and kept fixed for all experiments. For each class, 15\% of its subjects were held out for testing, including at least one subject from every severity level with two or more subjects. From these test subjects, 333 slices per class were drawn, spread as evenly as possible across subjects so that every brain has a similar weight in the test metrics. This selection depends only on a fixed random seed, so all models, training sizes and repetitions are evaluated on exactly the same test slices. The remaining subjects were divided into training and validation subjects using the seed of each repetition. At least one subject of every severity level was always kept for training. Validation subjects were drawn from the severity levels in proportion to their size until they provided 10\% of the training set size in slices per class or they made up at least 10\% of the subjects of the smallest class. The validation slices were then drawn evenly across these subjects. Finally, the training set was built by sampling the required number of slices per class (200, 500, 1000, 1500 and 3000) from the training subjects, again spread evenly across subjects. As a result, the three sets contain slices from disjoint groups of subjects, prevents subject-level data leakage. Only the training and validation subjects change between repetitions, the test set is the same in all of them, and no data augmentation was used. The exact subject distribution for training, validation and test are presented in Supplementary Information S6 for reproducibility.}

\subsection{Hyperparameter and quantum properties analysis}
\label{hyper}
This section provides a systematic investigation of how training hyperparameters and quantum properties influence the performance of the proposed hybrid quantum-inspired model and its classical counterparts for the balanced OCT dataset. We analyze each factor independently to identify its individual contribution, followed by a comprehensive comparison to elucidate the conditions under which the quantum-inspired approach provides \guille{better performance, in the studied datasets, than the considered} classical baselines. \guille{The best-performing hyperparameter configuration of each model, selected on the OCT validation set, is transferred to the OASIS dataset without further tuning.}

\subsubsection{Training size}

The amount of labeled data available for training is one of the most influential factors governing the performance of machine learning and quantum models. Since real world data acquisition and labeling can be costly or limited in practical settings, it is important to characterize how classification performance scales with the size of the training set, and whether classical and hybrid quantum-classical models exhibit similar or divergent learning behavior as more data becomes available. To this end, we systematically vary the training set size and evaluate its effect on \guille{validation} accuracy for all architectures under study.

Table \ref{tab:LRvsTRS} shows a relationship between classification performance, training set size, and learning rate. \guille{For the three models considered} (see Section \ref{sec:architectures}), \guille{ validation and test accuracy (see Supplementary Information S3)} generally improve as the training set grows from 200 to 30,000 samples, \guille{indicating} that additional training examples allow the architectures to learn more discriminative feature representations and \guille{generalize better}. This trend is particularly evident for learning rate 0.001, where \guille{accuracy increases almost monotonically with training set size}. The CNN\guille{-Dense} model reaches accuracies above 0.90 only when trained with the largest subsets, whereas considerably lower values are obtained when fewer than 1,000 samples are available. A similar pattern can be observed for HQiCNN, suggesting that both architectures benefit from the additional information contained in larger training sets. \guille{The CNN-MLP8 architecture yields lower accuracy than the HQiCNN for every training size and learning rate configuration considered. This is one of the central results of this study: since the models share the same convolutional backbone and the layer under study has exactly the same number of trainable parameters, the performance gap cannot be attributed to differences in model capacity. Instead, it indicates that, at equal parameter count, the emulated quantum layer provides a more expressive transformation than its classical counterpart. This is further supported by the results obtained for a CNN with a trigonometric MLP of eight parameters (see Supplementary Information S2), suggesting that this difference is not only found for ReLU-activated classical layers.}

\begin{figure}[h!]
    \centering
    \includegraphics[width=0.85\linewidth]{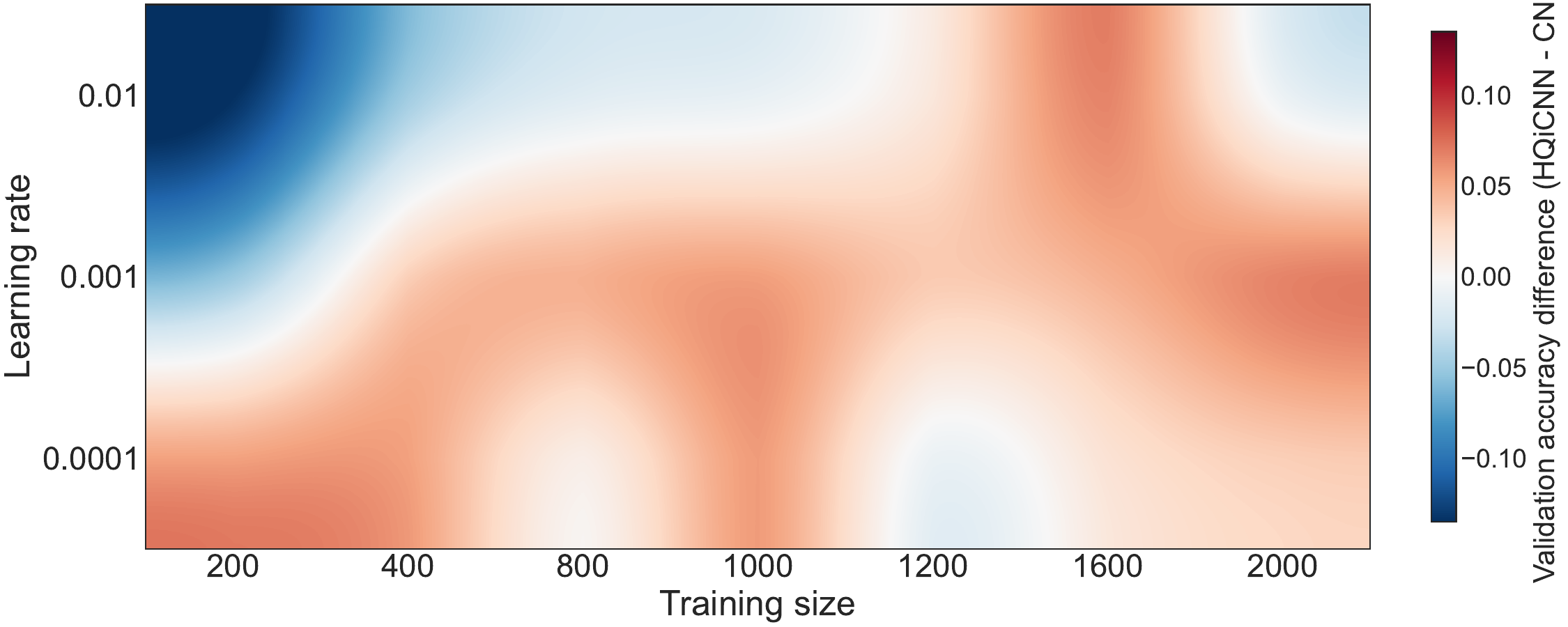}
    \caption{\guille{Validation accuracy difference between the HQiCNN and the CNN-Dense models. Red indicates greater performance of the HQiCNN and blue a better performance of the CNN-Dense model.}}
    \label{fig:val_LRvsTr_heatmap}
\end{figure}

Interestingly, the \guille{accuracy difference between the HQiCNN and the CNN-Dense architectures} is not constant across the explored range. The largest \guille{HQiCNN} gains occur when moving from very small datasets toward intermediate training sizes. Beyond approximately 12,000 samples, performance increases become more gradual, indicating that both models begin to approach a saturation regime in which additional data provide diminishing returns. This behavior suggests that the convolutional feature extractor is already capable of capturing most of the relevant information available in the dataset once a sufficient number of samples is provided. Although the overall trend is similar for \guille{the CNN-Dense and the HQiCNN} architectures, HQiCNN exhibits slightly higher performance in several intermediate training size configurations (see in Figure \ref{fig:val_LRvsTr_heatmap}).
% Interestingly, the magnitude of the improvement \guille{between the HQiCNN and the CNN-Dense architectures} is not constant across the explored range. The largest gains occur when moving from very small datasets toward intermediate training sizes. Beyond approximately 12,000 samples, performance increases become more gradual, indicating that both models begin to approach a saturation regime in which additional data provide diminishing returns. This behavior suggests that the convolutional feature extractor is already capable of capturing most of the relevant information available in the dataset once a sufficient number of samples is provided. Although the overall trend is similar for \guille{the CNN-Dense and the HQiCNN} architectures, HQiCNN exhibits slightly higher performance in several intermediate training size configurations (see in Figure \ref{fig:val_LRvsTr_heatmap}).
\begin{table}[t!] 
\centering
\small
\setlength{\tabcolsep}{3pt}
\renewcommand{\arraystretch}{1.1}
\caption{\guille{Mean validation accuracy (best epoch) over ten random seeds for the three models exploring different learning rates and training sizes. Values are reported as mean $\pm$ 95\% confidence interval half-width.}}\label{tab:valacc_lr_trsize}
\label{tab:LRvsTRS}
\begin{tabular*}{\textwidth}{@{\extracolsep{\fill}}lccccccccc@{}}
\toprule
\multicolumn{10}{c}{\textbf{CNN-MLP8}} \\
\midrule
\textbf{LR} & \multicolumn{9}{c}{\textbf{Training size}} \\
\cmidrule(lr){2-10}
 & $200$ & $400$ & $800$ & $1000$ & $1200$ & $1600$ & $2000$ & $12000$ & $30000$ \\
\midrule
$0.01$ & \acc{0.285}{0.068} & \acc{0.308}{0.076} & \acc{0.334}{0.100} & \acc{0.313}{0.096} & \acc{0.338}{0.102} & \acc{0.349}{0.114} & \acc{0.277}{0.061} & \acc{0.446}{0.181} & \acc{0.364}{0.172} \\
$0.001$ & \acc{\textbf{0.310}}{0.071} & \acc{\textbf{0.332}}{0.076} & \acc{0.374}{0.100} & \acc{0.383}{0.105} & \acc{\textbf{0.464}}{0.083} & \acc{\textbf{0.432}}{0.114} & \acc{\textbf{0.433}}{0.118} & \acc{\textbf{0.605}}{0.135} & \acc{0.561}{0.192} \\
$0.0001$ & \acc{0.270}{0.030} & \acc{0.265}{0.028} & \acc{\textbf{0.384}}{0.066} & \acc{\textbf{0.394}}{0.067} & \acc{0.416}{0.076} & \acc{0.374}{0.081} & \acc{0.417}{0.083} & \acc{0.541}{0.112} & \acc{\textbf{0.654}}{0.113} \\
\bottomrule
\end{tabular*}

\vspace{0.5cm}

\begin{tabular*}{\textwidth}{@{\extracolsep{\fill}}lccccccccc@{}}
\toprule
\multicolumn{10}{c}{\textbf{CNN-Dense}} \\
\midrule
\textbf{LR} & \multicolumn{9}{c}{\textbf{Training size}} \\
\cmidrule(lr){2-10}
 & $200$ & $400$ & $800$ & $1000$ & $1200$ & $1600$ & $2000$ & $12000$ & $30000$ \\
\midrule
$0.01$ & \acc{0.485}{0.075} & \acc{0.447}{0.121} & \acc{0.538}{0.050} & \acc{0.610}{0.059} & \acc{0.675}{0.091} & \acc{0.709}{0.068} & \acc{\textbf{0.809}}{0.017} & \acc{\textbf{0.900}}{0.005} & \acc{\textbf{0.920}}{0.004} \\
$0.001$ & \acc{\textbf{0.525}}{0.115} & \acc{\textbf{0.477}}{0.108} & \acc{\textbf{0.604}}{0.064} & \acc{\textbf{0.642}}{0.071} & \acc{\textbf{0.688}}{0.055} & \acc{\textbf{0.715}}{0.066} & \acc{0.736}{0.075} & \acc{0.896}{0.007} & \acc{0.919}{0.004} \\
$0.0001$ & \acc{0.280}{0.048} & \acc{0.335}{0.041} & \acc{0.521}{0.072} & \acc{0.488}{0.090} & \acc{0.588}{0.054} & \acc{0.614}{0.037} & \acc{0.625}{0.056} & \acc{0.841}{0.009} & \acc{0.884}{0.007} \\
\bottomrule
\end{tabular*}

\vspace{0.5cm}

\begin{tabular*}{\textwidth}{@{\extracolsep{\fill}}lccccccccc@{}}
\toprule
\multicolumn{10}{c}{\textbf{HQiCNN}} \\
\midrule
\textbf{LR} & \multicolumn{9}{c}{\textbf{Training size}} \\
\cmidrule(lr){2-10}
 & $200$ & $400$ & $800$ & $1000$ & $1200$ & $1600$ & $2000$ & $12000$ & $30000$ \\
\midrule
$0.01$ & \acc{0.415}{0.074} & \acc{0.405}{0.071} & \acc{0.514}{0.075} & \acc{0.602}{0.074} & \acc{0.680}{0.060} & \acc{\textbf{0.779}}{0.038} & \acc{0.794}{0.047} & \acc{\textbf{0.901}}{0.005} & \acc{0.918}{0.004} \\
$0.001$ & \acc{\textbf{0.460}}{0.108} & \acc{\textbf{0.520}}{0.079} & \acc{\textbf{0.631}}{0.074} & \acc{\textbf{0.700}}{0.040} & \acc{\textbf{0.728}}{0.041} & \acc{0.769}{0.036} & \acc{\textbf{0.807}}{0.030} & \acc{\textbf{0.901}}{0.005} & \acc{\textbf{0.923}}{0.003} \\
$0.0001$ & \acc{0.345}{0.064} & \acc{0.398}{0.071} & \acc{0.509}{0.065} & \acc{0.546}{0.054} & \acc{0.574}{0.046} & \acc{0.633}{0.024} & \acc{0.659}{0.031} & \acc{0.874}{0.011} & \acc{0.900}{0.006} \\
\bottomrule
\end{tabular*}
\end{table}

\subsubsection{Learning rate}

\guille{The learning rate has a substantial effect on optimization and, consequently, on the final classification performance. For all three architectures, a learning rate of 0.001 provides the best trade-off between accuracy and training stability. Increasing it to 0.01 returns worse results overall and makes the training less stable for the HQiCNN, as reflected in larger standard deviations and wider confidence intervals across seeds. Conversely, the lowest learning rate, 0.0001, systematically yields lower accuracies, which is consistent with slower convergence within the fixed number of training epochs.} This behavior suggests that the optimization process becomes excessively conservative, preventing the models from reaching highly discriminative solutions within the allocated training budget. Although performance still improves with increasing training size, \guille{no model} achieves the accuracy levels observed for larger learning rates. These findings indicate that the benefits of the hybrid quantum architecture are strongly linked to effective optimization. A poorly chosen learning rate can limit the ability of classical and quantum parameters to adapt to the training data, thereby masking potential \guille{improvements} associated with the quantum circuit \guille{architecture}.

To better understand the conditions under which HQiCNN provides \guille{better accuracy results} over CNN\guille{-Dense}, Figure \ref{fig:val_LRvsTr_heatmap} presents a heatmap showing the difference in mean \guille{validation} accuracy between \guille{the CNN-Dense and the HQiCNN} architectures in small and intermediate training size regimes (200-2000 images) \guille{, the corresponding test accuracy plot returns similar results and can be found in the Supplementary Information file (Sec. S3)}. Positive values indicate regions where HQiCNN outperforms CNN-Dense, whereas negative values indicate the opposite behavior. The heatmap reveals that the performance difference is not determined solely by training size or learning rate. Instead, it emerges from the interaction between both variables. Distinct regions of the parameter space exhibit different behaviors, suggesting that the effectiveness of the quantum layer depends on how optimization dynamics interact with data availability. \guille{We do not include the CNN-MLP8 results since the other models considerably outperform it across all tested configurations.} 

The most prominent positive region appears around intermediate training sizes \guille{($800-1600$ training samples)} and learning rates close to 0.001. In this area, HQiCNN consistently achieves higher accuracies than the classical baselines, indicating that the \guille{EQC embedded into a classical CNN architecture possibly provides a better interaction between optimization dynamics and data availability}. Negative regions are primarily concentrated at the extremes of the parameter space, particularly for very small datasets and for configurations where optimization is too aggressive. Under these conditions, the additional complexity introduced by the quantum circuit does not translate into improved predictive performance.

\subsubsection{Batch size}
\label{bs}
\begin{figure}[!t]
    \centering
    \begin{subfigure}{0.48\textwidth}
        \centering
        \includegraphics[width=\linewidth]{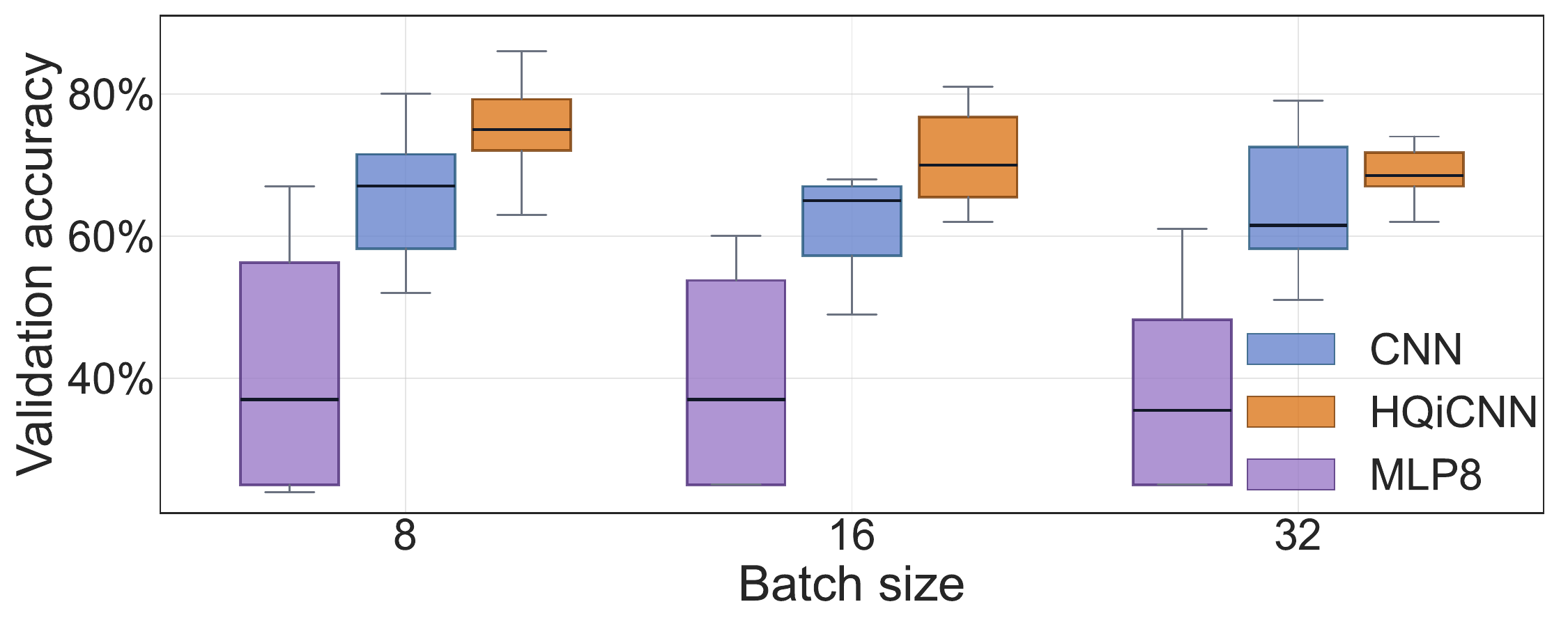}
        \caption{Accuracy comparison between the CNN and HQiCNN models for different batch sizes. \guille{This analysis is performed using a dataset of 1000 samples (see Section~\ref{bs})}}
        \label{fig:batchsize}
    \end{subfigure}
    \hfill
    \begin{subfigure}{0.48\textwidth}
        \centering
        \includegraphics[width=\linewidth]{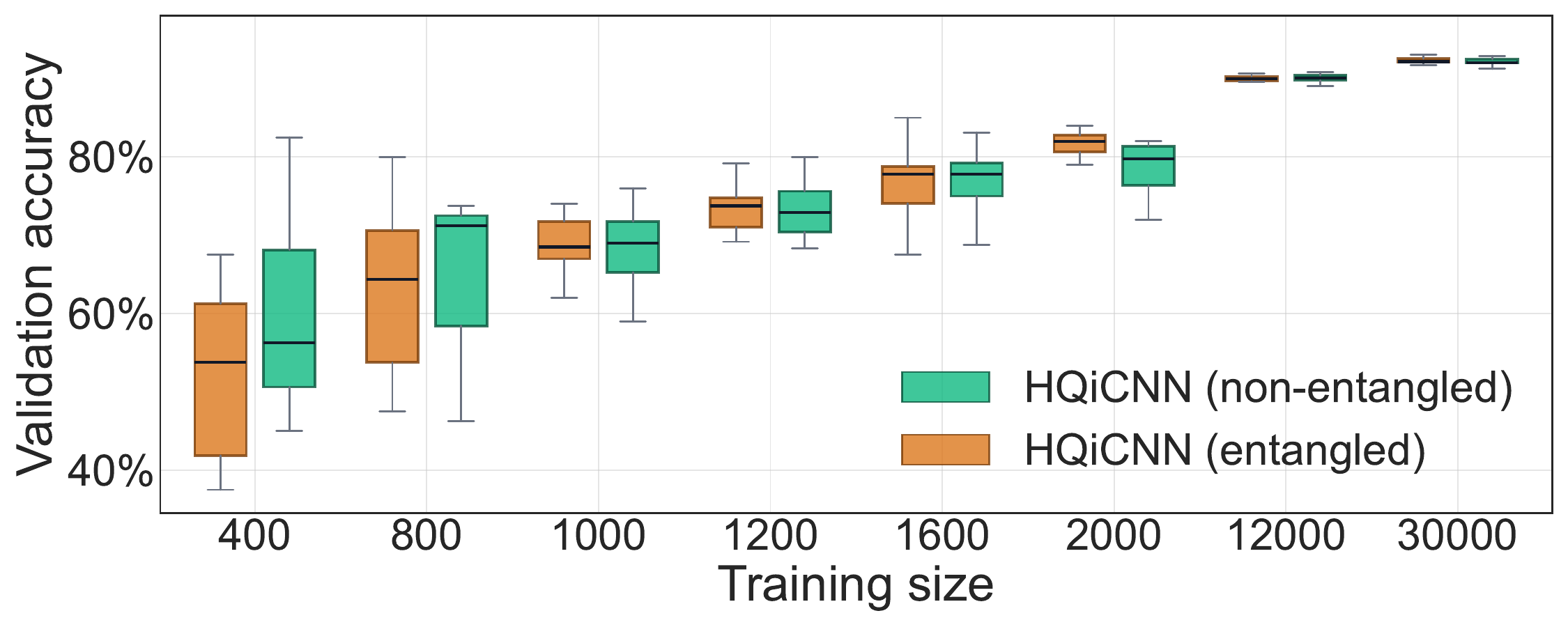}
        \caption{Accuracy comparison for the HQiCNN model with and without entanglement.}
        \label{fig:entvsnonent}
    \end{subfigure}
    \hfill
    \begin{subfigure}{0.48\textwidth}
        \centering
        \includegraphics[width=\linewidth]{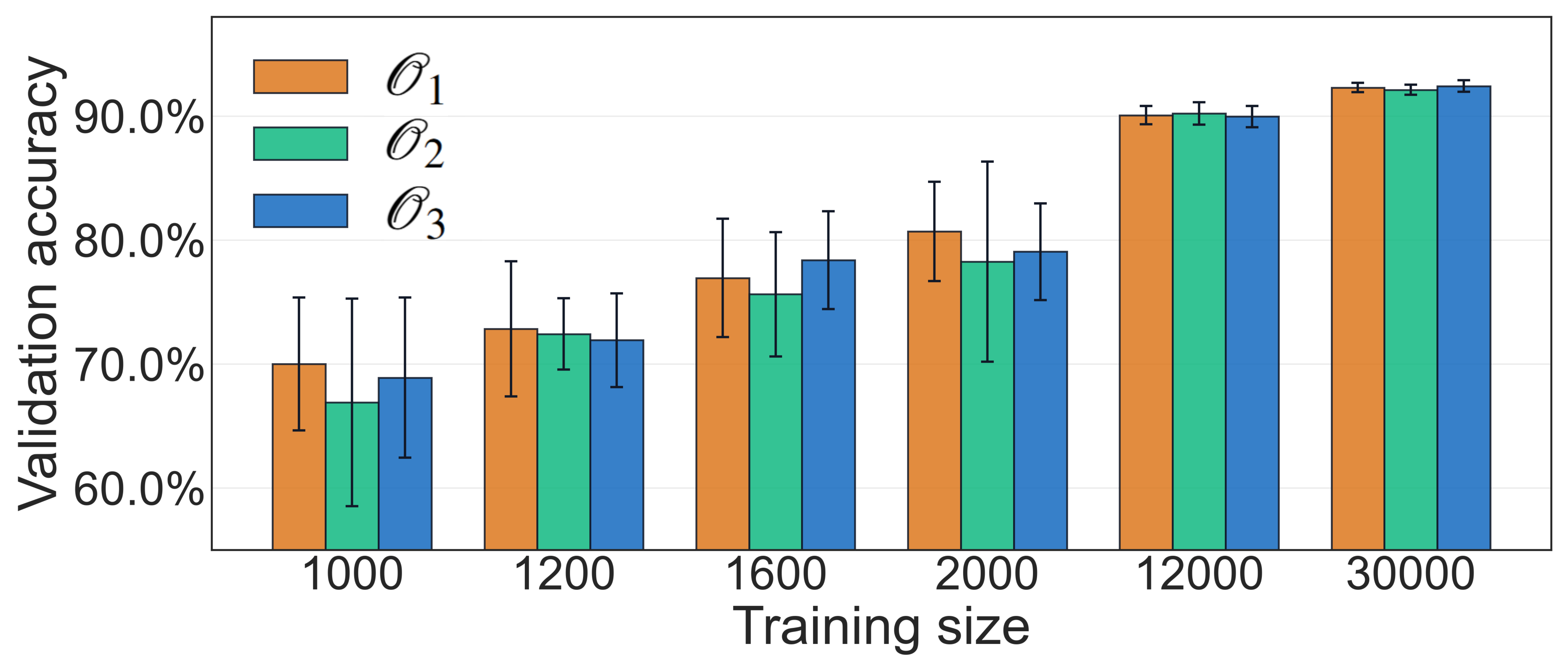}
        \caption{Accuracy comparison between the CNN and HQiCNN models for different \guille{observable sets.}}
        \label{fig:obsstudy}
    \end{subfigure}
    \caption{Batch size and quantum properties comparisons for the OCT dataset. In subfigures \ref{fig:bs_comparison_plot}\textcolor{blue}{a} and \ref{fig:entvsnonent}, the black horizontal lines of the box plots represent the median, the box covers from the first to the third quartile and the whiskers represent the maximum and minimum points within a $1.5\times IQR$ from the quartiles.}
    % \textcolor{blue}{All } }
    \label{fig:bs_comparison_plot}
\end{figure}

Batch size is a key hyperparameter in the training of neural network models, as it directly affects gradient estimation, convergence stability, and the effective noise present in the optimization process. In traditional sample-based hybrid quantum-classical models, batch size additionally influences on measurement statistics and the reliability of expectation-value estimates, making it particularly relevant to assess whether performance trends are robust across different batch size choices. Figure \ref{fig:batchsize} presents the performance of the models for a training size of 1000 samples across three different batch sizes. As has been extensively documented in the machine learning literature, hyperparameter optimization constitutes a critical factor in model performance. In this context, the aim of this analysis is to assess whether the observed difference in accuracy for the training size of 1000 remains consistent across variations in batch size. \guille{These statistical results, together with the half-width of the 95\% confidence interval, are available in Section S4 of the Supplementary Information.}\par
The quantum inspired hybrid model outperforms its classical counterparts in mean \guille{validation} accuracy, consistently across the three considered batch sizes. This result is consistent with previously published results regarding the optimization of Quantum Neural Networks hyperparameters \cite{moussa_hyperparameter_2024}, where they find that the batch size influences the runtime and memory but not the model performance.

\subsubsection{Entanglement}
The quantum circuit used in the previous results is the one shown in Figure \ref{fig:entvsnonent}, which includes an entanglement layer implemented via CNOT gates just before measurement. Note that the presence of entanglement is a widely used indicator of the need for a genuine quantum computer, as \guille{general} entangled circuits with non-Clifford gates are exponentially costly to simulate classically.\par

In Figure \ref{fig:entvsnonent}, we compare the results between the entangled and non-entangled versions of the circuit. On average, \guille{both settings perform similarly over all the training configurations; however, the entangled version usually returns more stable results for intermediate and large training size configurations (above $1000$ training samples, with the exception of $1600$ samples). We remark that our output observables are local and that they may not be able to fully capture the correlations generated by entanglement, potentially limiting the impact of entanglement in the classification results. \guille{The results included in Figure \ref{fig:entvsnonent}} can be found, with their half-width of the 95\% confidence interval, in Section S4 of the Supplementary Information.}

\guille{As observed in our experiments, HQiCNN configurations built from fully separable quantum circuits achieve performance comparable to their entangled counterparts. This indicates that, for the tasks considered here, entanglement is not required, and the HQiCNN model can therefore rely on separable circuits without a significant loss in accuracy.} \guille{The designed emulator uses a noiseless full statevector representation ($2^N$ amplitudes), but a separable circuit could instead be simulated with $N$ independent single-qubit states ($2N$ amplitudes), which would scale linearly in memory with the number of qubits, whereas general entangled circuits require exponential memory \cite{nielsen2010quantum}.}

\subsubsection{Observable set size}
\label{subsec:observable_set_size}
A crucial design choice in any quantum machine learning algorithm is the selection of the observable set used to extract classical information from the quantum state via measurement. This choice directly determines the dimensionality and expressivity of the feature space in which the subsequent classical model operates, and therefore has a direct bearing on predictive accuracy and computational cost. Since each additional non-commuting observable requires an independent expectation-value estimation on quantum hardware (or an additional shot budget in simulation), understanding how the size and structure of the observable set impacts performance is essential for balancing accuracy against resource requirements.

To investigate this trade-off, we compare three nested observable sets of increasing cardinality, corresponding to single-qubit (local), two-qubit (pairwise correlation), and three-qubit (triple correlation) Pauli terms, respectively. In Figure \ref{fig:obsstudy}, we directly compare the accuracy averaged over $10$ random seeds for $3$ different observable sets in the algorithm. These observable sets are:
\begin{itemize}
    \item $\mathcal{O}_1 = \{\Gamma_i\}\; \text{for}\; \Gamma \in \{X,Y,Z\} \;\text{and} \; i\in\{1,2,3,4\}$. Previous results were obtained using this observable set.
    \item $\mathcal{O}_2 = \mathcal{O}_1 \cup \{\Gamma_i \Gamma_j\}\; \text{for}\; \Gamma \in \{X,Y,Z\} \;\text{and} \; i,j\in\{1,2,3,4\} \; \text{with}\; i\neq j$.
    \item $\mathcal{O}_3 = \mathcal{O}_1 \cup \mathcal{O}_2 \cup \{\Gamma_i \Gamma_j\Gamma_k\}\; \text{for}\; \Gamma \in \{X,Y,Z\} \;\text{and} \; i,j,k\in\{1,2,3,4\} \; \text{with}\; i\neq j\neq k$.
\end{itemize}

These sets are nested, $\mathcal{O}_1 \subset \mathcal{O}_2 \subset \mathcal{O}_3$, so that any performance differences can be attributed purely to the additional higher-order correlators rather than to a change in the underlying single-qubit information already captured by $\mathcal{O}_1$.

\guille{For training set sizes below 2000 samples, the high variability of the results prevents us from identifying any clear trend with respect to the observable length. This suggests that, in this data regime, performance is limited primarily by the amount of training data rather than by the expressivity of the feature space. In contrast, for larger training sets (above $12000$ samples), the three observable sets converge to similar validation accuracies, and the observable length no longer affects performance. Hence, once sufficient training data is available, the higher-order correlators in $\mathcal{O}_2$ and $\mathcal{O}_3$ yield no measurable gain over local observables, indicating that the single-qubit observables in $\mathcal{O}_1$ already capture most of the discriminative information accessible from the EQC.} \guille{The mean validation accuracy for the results included in Figure \ref{fig:obsstudy} are available in Section S4 of the Supplementary Information, with their respective statistical  half-width 95\% confidence interval.}

% \guille{For training set sizes below 2000 samples, the high variability of the results prevents identifying any clear trend with respect to the observable length.} This suggests that in these data-regimes the model may be limited primarily by the amount of training data rather than by the expressivity of the feature space, and that the local single-qubit observables in $\mathcal{O}_1$ already capture the dominant part of the signal relevant for classification. However, large training sizes (above $12000$) show \guille{that the three considered observable sets converge to similar validation accuracy performance}, where the length of the observable \guille{does not affect the performance of the model}. This indicates that once sufficient training data is available, the additional higher-order correlators in $\mathcal{O}_2$ and $\mathcal{O}_3$ \guille{do not yield measurable gains over local observables alone, suggesting that $\mathcal{O}_1$ already captures most of the discriminative information accessible from the EQC.}

\guille{Thus, for the considered four-qubit EQC, local Pauli measurements already capture most of the information from the quantum layer, and the additional two-qubit and three-qubit  correlators are largely redundant for this task. Although this analysis is based on emulating circuits, the results imply that if the same circuits were run on quantum hardware, the extra measurement settings and shots required by $\mathcal{O}_2$ and $\mathcal{O}_3$ would add cost without a clear accuracy benefit, making $\mathcal{O}_1$ the more economical  and justified choice. Whether larger circuits or more demanding tasks would benefit from higher-order correlators remains an open question.}

\subsection{Comparative analysis between CNN and HQiCNN}
\guille{A direct comparison of the three architectures yields two main observations. First, the parameter-matched classical baseline (CNN-MLP8) performs substantially worse than the HQiCNN, whose EQC contains the same number of trainable parameters. This suggests that, under a strict parameter budget, the EQC acts as a more expressive feature-processing block than a small classical MLP, enabling the HQiCNN to extract more useful information per trainable parameter. This is further validated for an alternative trigonometric MLP with the same number of parameters in the Supplementary Information S2. Second, the CNN-Dense model, a stronger classical baseline, performs comparably to the HQiCNN, and neither model consistently outperforms the other across the tested conditions. Instead, their relative performance depends on the interplay between training set size and learning rate. The modest gains of the HQiCNN at intermediate training set sizes are empirical observations obtained through classical emulation of quantum circuits, and should not be interpreted as evidence of quantum advantage.}

\guille{The CNN-Dense maps the reduced latent representation to the output through a dense layer, whereas the HQiCNN first encodes these latent features into a four-qubit parameterized EQC, whose measured outputs are then passed to the final classification layer (see Figure~\ref{fig:architectures}). Since both models share the same convolutional backbone, the slightly better performance of the HQiCNN in certain regimes can be attributed primarily to the replacement of the dense layer by the parameterized EQC.} However, the limited magnitude of the observed gains also suggests that the shared convolutional backbone already extracts the majority of the discriminative information, so the \guille{emulated} quantum layer acts primarily as a refinement mechanism rather than as a complete replacement for classical feature learning. As a result, we find that the relatively small performance gap between the two approaches is itself an important finding, it indicates that hybrid quantum-inspired architectures can achieve results comparable to \guille{their classical baseline}, supporting the feasibility of integrating \guille{emulated} quantum layers into practical machine-learning pipelines.
\subsection{Cross-Dataset Robustness: Alzheimer's Disease Classification}

In previous subsections, we have studied the performance of the models under different hyperparameter configurations using the OCT dataset, a large and highly balanced dataset that enabled a comprehensive evaluation across a wide range of experimental settings. In this subsection, we transfer the best-performing hyperparameter configuration obtained from the OCT experiments to the OASIS dementia MRI dataset (see Section \ref{sec:datasets}) to evaluate the robustness and generalization capability of the proposed framework in a different medical imaging scenario. \guille{We remark that the evaluation is at the slice label and that the transferred hyperparameters were selected based on validation accuracy, limiting potential test-set bias in model selection.} \guille{To obtain a balanced binary problem, the original OASIS categories were grouped into Non Demented and Demented, the latter combining the Very Mild, Mild and Moderate Dementia classes. This reduces the class imbalance of the original dataset while keeping a meaningful distinction between healthy and demented subjects (see Sections \ref{sec:OASIS_intro} and \ref{trteva} for details).}
% \deleted{To obtain a balanced and clinically meaningful classification problem, the original OASIS categories were reorganized into \guille{two classes: Non Demented and Demented. The latter is generated by combining the Very Mild Dementia, Mild Dementia and Moderate Dementia classes (See Figure \ref{fig:dementia_plot})}. This reformulation allows the model to distinguish between cognitively healthy subjects and \guille{demented ones}, while reducing the class imbalance present in the original dataset \guille{(See Section 4.1.1 for a more comprehensive explanation)}}.

\begin{figure}[h!]
\centering
\includegraphics[width=\linewidth]{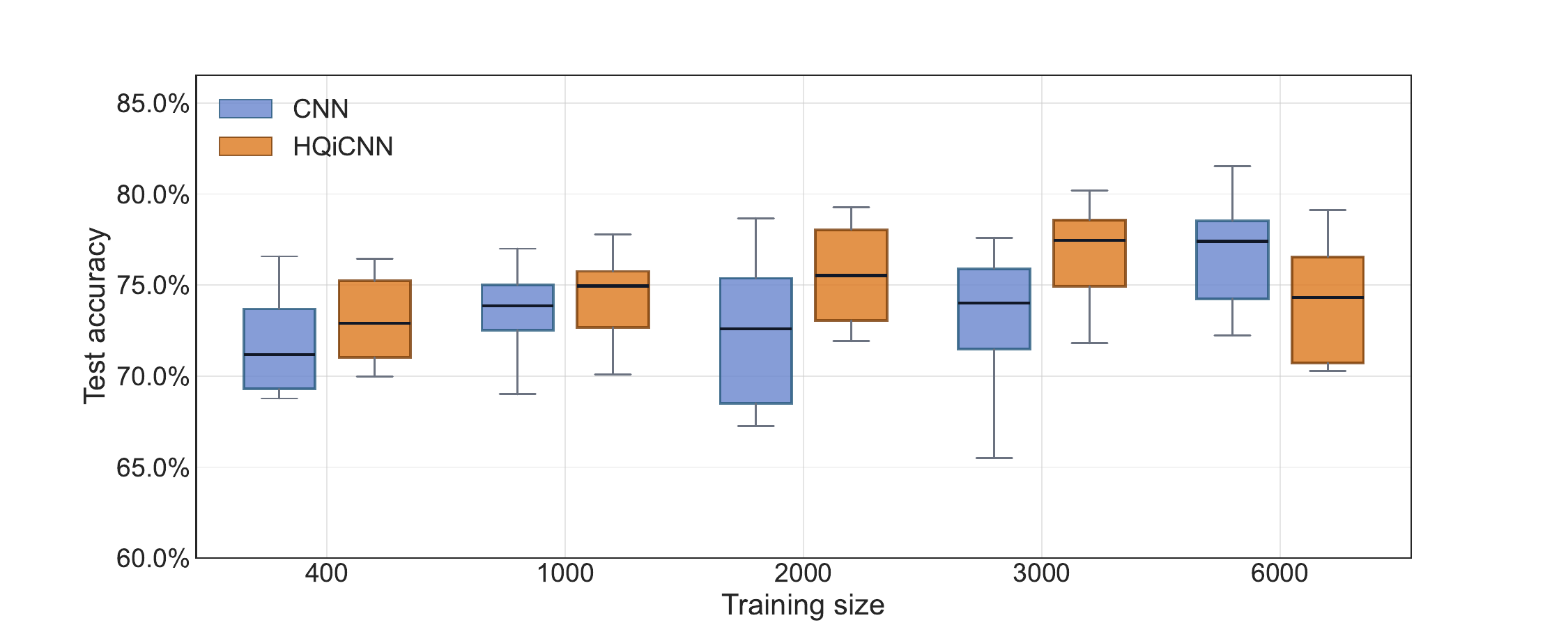}
\caption{Accuracy comparison between the HQiCNN and CNN models for the dementia dataset (see Section 4.1). Each model uses the best learning rate found for the most comparable training size in Table \ref{tab:LRvsTRS}. The black horizontal lines of the box plots represent the median, the box covers from the first to the third quartile and the whiskers represent the maximum and minimum points within a $1.5\times IQR$ from the quartiles.}
\label{fig:dementia_plot}
\end{figure}
Figure \ref{fig:dementia_plot} shows the \guille{test}  accuracy results for different training sizes. We find \guille{that these results are consistent with }the OCT results shown in Table \ref{tab:LRvsTRS}\guille{, especially in the intermediate training size regime (1000-3000 training samples); } thus, the \guille{relative behavior of the CNN-Dense and HQiCNN is preserved for} two different health-related image classification tasks \guille{with similar improvements observed in comparable training set size regimes}. \guille{However, when the training set is increased to $6000$ samples, the performance of the HQiCNN decreases, whereas the accuracy of the CNN-Dense continues to improve. This suggests that the results are consistent with those obtained for OCT in the small and intermediate training-size regimes, while for larger training sets the HQiCNN is outperformed by the CNN-Dense, although they remain competitive over the tested training regimes.}

\subsection{Explainability}
\guille{We apply SHAP (SHapley Additive exPlanations)~\cite{lundberg2017unified} to the OCT images in order to analyze the decision-making process of the proposed models. Note that this analysis is not intended to validate the clinical relevance of the highlighted regions, which would require assessment by ophthalmology experts. Instead, it focuses on comparing the behavior of HQiCNN with that of a classical CNN, a well-established and widely validated framework for machine learning in healthcare. Since the HQiCNN and the CNN-Dense architectures are identical except for a single layer, any differences observed in their attribution maps can be directly attributed to the quantum component. In this way, SHAP provides a controlled means of assessing how the quantum layer influences the features the model relies on for classification.}
% To further investigate the decision-making process of the proposed models, a SHAP  \cite{lundberg2017unified} based tool analysis was performed on OCT images.\guille{This explainability study should not be seen as a medical study, rather as a explainable comparative research between the model we are testing (HQiCNN) and a commonly used and validated framework in ML for health purposes (CNNs).}

\guille{We first study the $|\text{SHAP}|_{\text{IoU}}$ metric for the top 2\% and top 10\% ranked values, that illustrates} the spatial overlap of the \guille{top corresponding percentages of the} absolute SHAP values for HQiCNN and a conventional CNN. Red regions correspond to image areas exclusively identified as relevant by HQiCNN, blue regions indicate areas uniquely emphasized by the CNN\guille{-Dense}, and green regions represent the regions jointly considered important by both models. \guille{As explained in \ref{sec:SHAPiou_explain}, this metric aggregates over seeds, limiting possible seed bias. The two thresholds probe agreement at two spatial scales: the top 2\% isolates the small set of pixels that dominate each attribution map, while the top 10\% covers the broader region supporting the prediction. More permissive thresholds (20\% and 60\%) are reported in Supplementary Section S5; at these levels the selected regions cover large parts of the image and the overlap between models no longer discriminates model-specific attention.} 
\begin{figure}[!t]
    \centering
    \begin{subfigure}{0.9\textwidth}
        \centering
        \includegraphics[width=\linewidth]{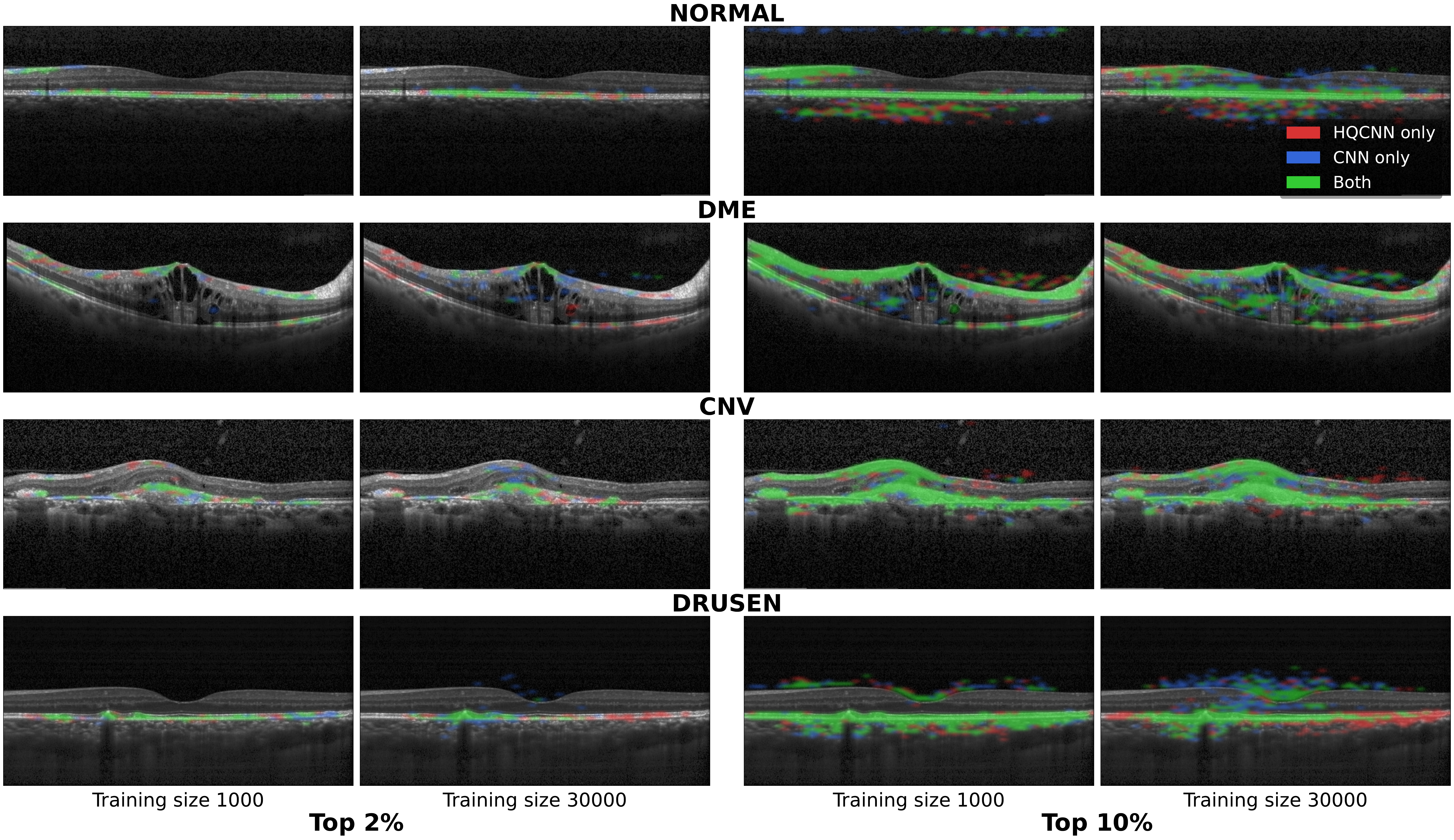}
        \caption{$|\text{SHAP}|_{\text{IoU}}$ across four lables, two training sizes and top 2\% adn 10\% ranked $|\text{SHAP}|_\text{IoU}$. Red regions denote the areas of highest influence for the HQiCNN model’s classification decision, blue regions denote the corresponding areas for the CNN\guille{-Dense} model, and green regions represent the intersection of influence between the two models.}
        \label{fig:shapiou}
    \end{subfigure}
    \hfill
    \begin{subfigure}{0.7\textwidth}
        \centering
        \includegraphics[width=\linewidth]{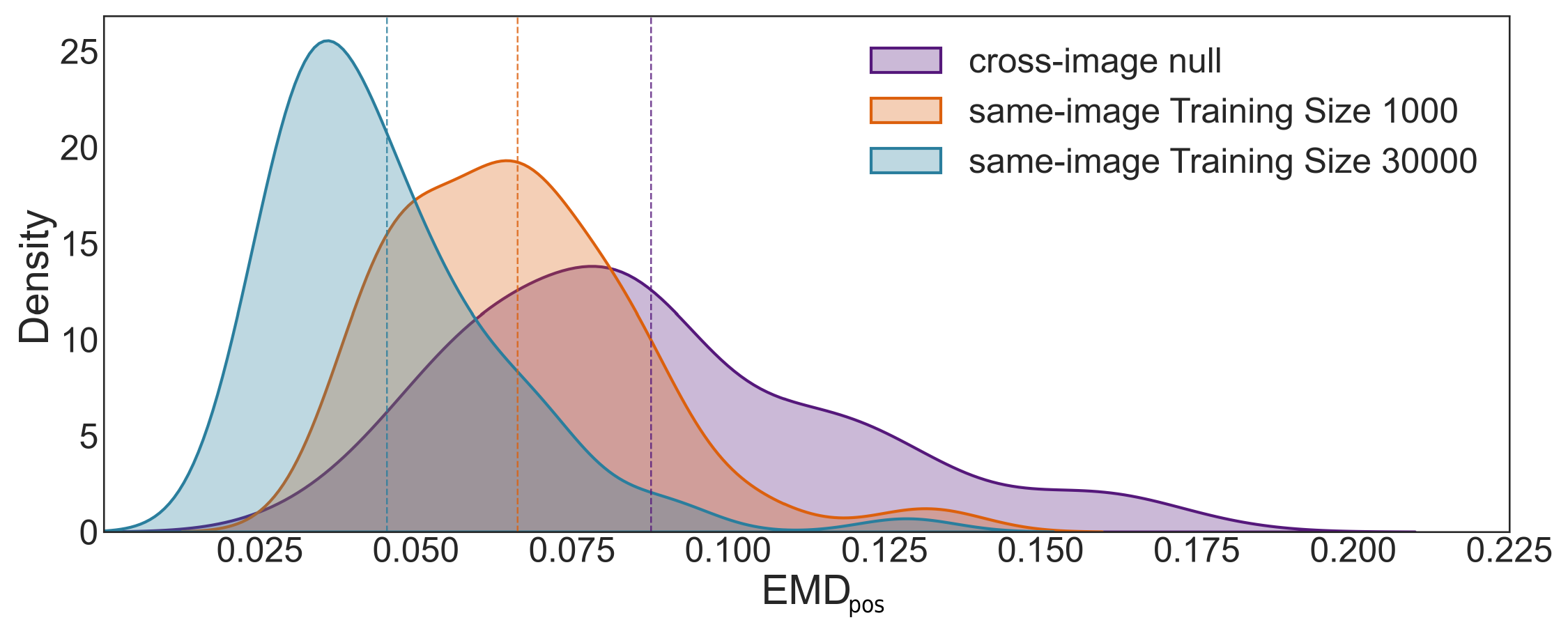}
        \caption{Image count distribution against the $\text{EMD}_\text{pos}$  value for $1000$ and $30000$ training sizes on the same images and the combined different image setting. The dashed vertical lines indicate the mean for each distribution.}
        \label{fig:emdpos}
    \end{subfigure}
    \caption{Explainability plots for comparing \guille{the HQiCNN and the CNN-Dense} models predictions for the OCT dataset.}
    \label{fig:shap_all_oct}
\end{figure}

On one hand, the top 2\% ranked results show that both architectures focus primarily on the retinal regions affected by the three considered diseases, particularly centered on the characteristic \guille{retinal areas where lesions occur} (see Section \ref{sec:datasets}), confirming that their predictions are based on \guille{areas that were identified as regions of clinical interest for the diagnosis of these diseases in other studies that used the same dataset and that were validated by clinicians~\cite{kermany2018identifying}}. \guille{This happens for the considered training sizes, indicating that the two models pay the most attention, even under different training size regimes, to the same areas of plausible anatomical interest. On the other hand, the top 10\% ranked results show that both models still attend to the clinically interesting areas. However, we can see some deviations in the CNN-Dense and the HQiCNN. For the experiment with 1000 training samples, the CNN attends to regions close to the image borders far from the anatomically plausible areas in the NORMAL image, while the HQiCNN heatmaps are more concentrated around them. However, with increased training data this behavior is no longer observed, with the CNN concentrating its attention on anatomically plausible regions.}

%\guille{Thus, the $|\text{SHAP}|_\text{IoU}$ metric shows that both models place their highest attributions (top $2\%$ ranked) on the clinically relevant retinal regions identified by clinicians~\cite{kermany2018identifying}. However, when lower-importance regions are included (top $10\%$ ranked), the CNN under small training sets shifts part of its attribution towards the image borders.}

In order to obtain an averaged measure of similarity between the \guille{positive} SHAP distributions, and to avoid relying exclusively on single-image interpretations, we use the previously introduced $\text{EMD}_\text{pos}$ metric. This measure enables us to quantify the \guille{spatial} similarity between the positive SHAP distributions of the studied models over a set of considered images. We obtain $0.066 \pm 0.019$ for the models trained with $1000$ samples and $0.045 \pm 0.018$ for those trained with $30000$ samples. \guille{Since the pixel coordinates are normalized by the image side length, $\text{EMD}_\text{pos}$ is expressed in units of image width. Multiplying by $64$ therefore recovers the average displacement, in pixels, required to transport one positive-SHAP distribution onto the other. Both configurations correspond to $\sim 4.2$ and $\sim 2.9$ pixels, respectively.} \guille{Increasing the training size thus reduces the spatial separation between the HQiCNN and CNN positive SHAP distributions to a few pixels on a $64 \times 64$ image.}

\guille{To provide an empirical reference for these magnitudes, we compare them with a cross-image null distribution (see Fig. \ref{fig:emdpos}), in which $\text{EMD}_\text{pos}$ is computed between SHAP maps of different images (see Section~\ref{sec:emdpos}). This reference is wider, with a mean of $0.088 \pm 0.043$, equivalent to an average displacement of $\sim 5.7$ pixels. In Figure~\ref{fig:emdpos}, the $1000$-sample configuration has a lower mean but overlaps considerably with the cross-image distribution, which is expected if the anatomy is aligned across images. When the training size increases to $30000$, the mean $\text{EMD}_\text{pos}$ decreases further and the overlap with the null shrinks, indicating that the spatial similarity between the CNN-Dense and the HQiCNN increases with training size.}

\guille{This combined analysis shows that, across different training data regimes, the HQiCNN consistently reproduces the most important attributions of the classical CNN architectures for the same images, while its positive SHAP explanations remain spatially close to theirs over the entire test set. These findings strengthen the comparison between the models, since the similarity of their explanations provides additional, explainability-based support for the accuracy results reported above. Note, however, that this analysis characterizes the behavior of the models, namely which input regions drive their predictions and how consistently they do so, rather than the relevance of these regions for the classification task itself.}

\section{Conclusions and future work}
\label{sec:conclusions}
This work presents small emulated quantum circuits as a tool for improving the performance of standard networks, and a systematic evaluation \guille{of whether and under which conditions the HQiCNN outperforms comparable classical baselines} on real world medical image classification tasks. By keeping the convolutional backbone identical and replacing only an intermediate layer, \guille{we study how a classically emulated quantum layer embedded in a classical CNN architecture performs under different experimental configurations and compare it with two classical alternatives.}

\guille{The experimental results show that, under the same parameter budget, the HQiCNN considerably outperforms the parameter-matched classical baseline (CNN-MLP8) under ReLU and sinusoidal (see Supplementary Information S2) activation functions. Moreover, the HQiCNN remains competitive with the CNN-Dense, a standard classical configuration with approximately six times more trainable parameters. The HQiCNN achieves the highest overall accuracy on the OCT dataset (up to 92.3\%) and performs best in the intermediate training size regimes, whereas the CNN-Dense remains competitive overall, particularly above 2000 training samples.}

The systematic hyperparameter analysis further shows that the effectiveness of the HQiCNN \guille{and the CNN-Dense} depends on the interaction between data availability and optimization. Empirically, we show that a learning rate of $10^{-3}$ provides the most stable performance \guille{overall, and the three tested models maintain consistent results for different batch sizes.}  \guille{Furthermore,} removing entanglement produces comparable, and often slightly better, classification accuracy. This finding is particularly relevant because separable circuits scale linearly \guille{, while entangled circuits scale exponentially in memory for exact statevector simulation.} \guille{Extending the observable set beyond local Pauli measurements ($\mathcal{O}_1$)  to include two-body ($\mathcal{O}_2$) and three-body ($\mathcal{O}_3$) correlators yields} \guille{different results depending on the training size. At lower training sizes, we find large variability with no clear trend between observable sets. At larger training sizes, the three sets converge to nearly identical results. These results suggest that local Pauli measurements ($\mathcal{O}_1$) capture enough information from the emulated quantum circuit, so the additional measurement cost of higher-order correlators ($\mathcal{O}_2$, $\mathcal{O}_3$) offers no clear benefit in the regimes explored.}

From an interpretability perspective, \guille{we have compared the HQiCNN and the CNN-Dense models using the proposed} $|$SHAP$|_{\text{IoU}}$ and EMD$_{\mathrm{pos}}$ metrics. The  $|\text{SHAP}|_\text{IoU}$ shows \guille{that the two considered architectures primarily focus on the same regions (top 2\% ranked), located in anatomically plausible areas, for the two training settings considered.} \guille{However, other lower importance areas of attention (top 10\% ranked) show that both models start to deviate from these areas, especially the CNN with limited training data}. The average EMD$_{\mathrm{pos}}$ decreases from $0.066 \pm 0.019$ for $1000$ training samples to $0.045 \pm 0.018$ for $30000$ samples, indicating a \guille{trend toward closer} agreement between the models as more data becomes available. \guille{Thus, the comparative explainability analysis shows convergence between HQiCNN and CNN's attribution maps, indicating that the emulated quantum layer does not substantially alter where the classical CNN-Dense network focuses.}

\guille{These} results indicate that \guille{the HQiCNN and CNN-Dense architectures perform similarly, with the HQiCNN providing modest improvements at intermediate training set sizes ($800$ and $1600$ samples). 
Note that these relatively small improvements, and the fact that the HQiCNN architecture should be regarded as a quantum-inspired model, with no intended execution on real quantum hardware, provide no evidence of a possible quantum advantage.} \guille{This inspired approach for four-qubit EQCs allows for fast running on classical computers and avoids the challenges associated with practical quantum computation, so these results should not be  extrapolated to quantum hardware.} 
% \guille{These effects may alter the behavior and scalability of such models on real devices, so our conclusions should not be extrapolated to quantum hardware.}
% presented approach is based on emulated shallow circuits with four qubits, which allows f In particular, these models do not face any additional limitations imposed by the quantum hardware, such as noise accumulation, decoherence effects, statistical uncertainty from finite-shot measurements, and restrictions imposed by the physical connectivity of quantum processors
% The experiments provide no evidence of quantum advantage in the HQiCNN architecture and suggest that the improvements observed in these regimes arise from the interplay between the amount of training data, the optimization strategy and the chosen measurement scheme.} 

%However, the current analysis is based on emulated shallow circuits with four qubits, which allows a controlled comparison of different configurations but does not fully represent the challenges associated with practical quantum computation. In particular, realistic NISQ devices introduce additional limitations, such as noise accumulation, decoherence effects, statistical uncertainty from finite-shot measurements, and restrictions imposed by the physical connectivity of quantum processors, which may affect the behavior and scalability of these models.

\guille{Despite these promising results, several aspects of this work warrant further investigation. Future work will extend the evaluation to additional cohorts and multicenter datasets acquired with different scanners and imaging protocols, in order to assess the robustness and the generalization of the proposed framework across heterogeneous populations and acquisition conditions. In addition, studies involving clinicians would help assess the diagnostic utility and interpretability of the model’s predictions in real clinical settings. The use of a 2D slice-based representation of the MRI volumes, rather than their full 3D structure, may limit the anatomical context available to the model. Although predictive performance was evaluated comprehensively across multiple experimental settings, the associated computational cost, including training time, inference efficiency and memory requirements, was not analyzed. Future work will address these limitations by validating the proposed framework and extending it to 3D representations, while analyzing its computational efficiency in detail. In addition, we plan to explore larger emulable quantum circuits and other quantum-inspired techniques, such as tensor networks and Pauli propagation, in order to characterize the trade-offs between predictive performance and resource requirements in hybrid quantum-inspired learning models.}

% \textcolor{blue}{Adding here limitations too, and future lines (measuring memory and timing in empty GPU etc...) }
\newpage

\section*{Data availability}
The data supporting the experiments conducted in this article are the OCT scans, available in Reference \cite{kermany2018oct}, and the OASIS1 images dataset \cite{marcus2007oasis} used for the technical robustness section, available at \url{https://www.kaggle.com/datasets/ninadaithal/imagesoasis}. Details related to the OASIS MRI dataset preprocessing can be found at \url{https://www.kaggle.com/datasets/ninadaithal/oasis-1-shinohara/}.

\section*{Code availability}
The code used in this study can be made available from the corresponding authors upon reasonable request.

\bibliography{sample}

\section*{Funding}

This article is funded by PESL-Stiftung-Alzheimer 2026 in Bayern, Germany, (registered project-80765114), which is focused on research in Alzheimer’s disease. Furthermore, researchers G.R., M.O., M.A., G.B. and P.D. acknowledge the support provided by project ARQADE (CER-20251019), funded by the CERVERA Research Programme of CDTI (Centre for Technological Development and Innovation).

\section*{Author contributions statement}
G.R., M.O., M.A., P.D., and S.A. conceived and planned the experiments. G.R. and M.O. carried out the experiments. G.B.,  S.A., and B.R. provided funding. G.R., M.O., M.A., P.D., and S.A. contributed to the interpretation of the results. G.R. and S.A. took the lead in writing the manuscript. 
\vspace{2mm}

\noindent All authors provided critical feedback and helped shape the research, analysis and manuscript.

%https://www.elsevier.com/researcher/author/policies-and-guidelines/credit-author-statement
%  conceptualization, C.A., G.V., and L.R-C; validation, L.R-C., J.R. and S.B.; formal analysis, G.C.; investigation, G.C.; writing—original draft preparation, J.R., L.R-C., and C.A.; writing—revision and editing, G.V., S.B.,  C.A., and J.R.; supervision, L.R-C. and S.B.;. All authors have read and agreed to the published version of the manuscript
%  EXAMPLE
%A.B., B.C., C.D., D.E., E.F., F.G., and G.H. conceived and planned the experiments. A.B., B.C., C.D. and D.E. carried out the experiments. A.B., F.G. and E.F. planned and carried out the simulations. J.K., K.L., A.B., B.C., D.E., C.D., F.J., and F.G. contributed to sample preparation. A.B., B.C., C.D., D.E., FJ, E.F., F.G. and G.H. contributed to the interpretation of the results. A.B. took the lead in writing the manuscript. All authors provided critical feedback and helped shape the research, analysis and manuscript. 

\section*{Additional information}

\textbf{Competing financial interests:} The authors declare no competing financial interests.

% To include, in this order: \textbf{Accession codes} (where applicable); \textbf{Competing interests} (mandatory statement). 

% The corresponding author is responsible for submitting a \href{http://www.nature.com/srep/policies/index.html#competing}{competing interests statement} on behalf of all authors of the paper. This statement must be included in the submitted article file.
\newpage
\end{document}